\documentclass[%
 aps,
 amsmath,amssymb,
preprint,
]{revtex4-1}

\usepackage{graphicx}
\usepackage{dcolumn}
\usepackage{bm}

\usepackage[utf8]{inputenc}
\usepackage[T1]{fontenc}
\usepackage{mathptmx}
\usepackage{etoolbox}
\usepackage{subcaption}
\usepackage{physics}

\begin{document}


\title[First-Principles Calculation of Alloy Scattering and n-type Mobility in Strained GeSn]{First-Principles Calculation of Alloy Scattering and n-type Mobility in Strained GeSn}

\author{K. Sewell}
\author{F. Murphy-Armando}
 \email{philip.murphy@tyndall.ie}
\affiliation{ 
Tyndall National Institute, University College Cork, T12R5CP, Ireland.
}

\date{\today}

\begin{abstract}
We use first-principles electronic-structure theory to determine the intra- and inter-valley electron-alloy scattering parameters in n-type GeSn alloys. These parameters are used to determine the alloy scattering contributions to the n-type electron mobility of GeSn at $300K$ and $15K$ using a first iteration of the Boltzmann transport equation in the relaxation time approximation. For unstrained GeSn, we find that a Sn concentration of at least $13.5\%$ is needed to achieve an electron mobility greater than that of Ge. Our results show that the mobility of GeSn can be over $25$ times higher than the mobility of Ge, or $10^5$ cm$^2$/(Vs). At $15K$, less than $6\%$ Sn incorporation into Ge quadruples its mobility, which suggests GeSn has potential applications as a high mobility 2D electron gas. Applying biaxial tensile strain to GeSn further increases the mobility and at a lower Sn content than in unstrained GeSn.
\end{abstract}

\maketitle

\section{\label{sec:intro}Introduction}

GeSn alloys are increasingly of interest in a range of applications. In optoelectronics, GeSn has been demonstrated in light-emitting diodes and photodiodes \cite{cardoux2022room,huang2019electrically,chang2017room}, in lasers \cite{kim2022improved,ojo2022silicon,marzban2022strain,wirths2015lasing}, photo-detectors \cite{atalla2022high,talamas2021cmos,tran2019si,zhang2018gesn}, in optical interconnects \cite{miller2009device,liu2018chip} and lab-on-chip trace gas detection \cite{lavchiev2016photonics}, as well as becoming a promising candidate to enable monolithic on-chip Si photonics operating in the near-infrared and short-wave infrared wavelengths \cite{dutt,liu2007tensile,wirths2013tensely,giunto2023gesn,moutanabbir2021monolithic}.
GeSn has also been previously suggested as a high mobility material \cite{sau,mukh}, and experimental measurements of the electron mobility with tensile strain has prompted research for GeSn FETs \cite{chuang2020electron,huang2017high}. On the other hand, strained Ge has been proposed as a high hole-mobility material, which has resulted in record-high 2D hole gas mobilities in strained-Ge QW modulation-doped heterostructures \cite{scappucci2021germanium,melnikov2015ultra}. In addition, Myronov et al \cite{myronov2023holes} have measured a record-high hole mobility of $4.3 \times 10^6 \ cm^2/(Vs)$ in strained Ge at $300mK$, which has fueled the development of quantum-electronic circuits based on spin-qubits. Currently, there are some issues associated with GeSn processing. Due to the lattice mismatch between Ge and Sn, the solubility of Sn in Ge is low \cite{ayan2019strain,thurmond1953equilibrium}. Sn tends to segregate out of the Ge crystal without out-of-equilibrium processing, resulting in the formation of $\beta$-Sn, which is detrimental for opto-electronic devices \cite{li2013strain,cai2022thickness,tsukamoto2015investigation,li2014characteristics,wang2020effects}. At present, as much as $18\%$ Sn has been successfully incorporated into Ge \cite{kong2022growth}.

The appeal of GeSn largely stems from its tunable bandgap nature. Ge, which has a relatively high electron mobility of $\mu\simeq\frac{e}{m^*R}\approx3900 \ cm^2/(Vs)$ \cite{Fisch,prince} (where $R$ is the scattering rate), and has a pseudo-direct-bandgap, whereby a difference of around $150 \ meV$ separates the lowest $L-$ and higher $\Gamma -$ conduction band minima \cite{Fisch,rideau,kurdi,kao}. Adding Sn to Ge achieves an indirect-to-direct transition at Sn compositions 7-9\% Sn concentration \cite{jiang,ghetmiri2014direct,gallagher2014compositional}. Direct-gap GeSn is desirable since the $\Gamma$ valley usually has the lightest effective mass $m^*$ in diamond and zinc-blende semiconductors such as GeSn or GaAs, resulting in a high electron mobility. The $\Gamma$ valley is also optically coupled to the valence band, where radiative recombination is much faster, leading to improved optical properties. Another method of inducing a direct-gap in Ge is to introduce biaxial tensile strain
, which lowers the $\Gamma$ valley below the $L$ valley, and strain $\geq 1.5\%$  is enough to produce a direct-gap in Ge \cite{kurdi,kao,clavelfma}. Menendez et al \cite{soref2006advances} have predicted that tensile-strained Ge exceeding $2\%$ is achievable before breakage, which has been achieved in Refs \cite{saladukha2018direct,pavarelli2013optical}.

In this work, in order to minimise Sn segregation and to avoid strain much above the critical thickness, we propose a combination of strain and Sn alloying to increase the intrinsic, n-type electron mobility of GeSn. However, the advantage of achieving a direct gap by alloying Ge with Sn maybe be offset by the introduction of alloy scattering, which diminishes the electron mobility \cite{Fisch}. The alloy scattering parameters for GeSn have not been previously calculated theoretically for GeSn, and are difficult to determine from experimentally measured quantities \cite{fma06} due to Sn segregation. Sau and Cohen \cite{sau} have used a first-principles method developed by Murphy-Armando and Fahy \cite{fma06} to calculate a single intra-valley scattering matrix element using both an eigenvalue shift method and a method of splitting of degeneracies at the edge of the folded-back supercell Brillouin Zone. Both methods present complications: the eigenvalue shift method requires the additional calculation of the lattice potential, and using the splitting approach can lead to significant errors for the extremely light conduction band state at $\Gamma$ in a $128-$atom supercell such as the one they used. However, these values should converge in large supercells. They report values of the intra-valley scattering at $\Gamma$ ($0.788 \ eV$ and $0.1 \ eV$) and predict an electron mobility of $Ge_{0.92}Sn_{0.08}$ of order $10^6$. However, given that both intra- and inter-valley alloy scattering to and from the $L$ valley weren't included in their calculation, this is likely to be an overestimation, especially given that the $L$-mobility is much more dominant than the $\Gamma$-mobility at $8\%$ Sn due to its 33-times larger density of states. We employ the same first-principles methodologies as \cite{fma06} to ascertain the intra- and inter-valley alloy scattering parameters in n-type $Ge_{1-x}Sn_x$. The derived properties are then used to determine the alloy scattering contributions to the n-type electron mobility in GeSn, which is calculated as a function of alloy composition, biaxial strain and temperature. We perform calculations at room temperature ($300K$) and at a lower temperature of $15K$, since GeSn is a candidate material in quantum computation \cite{myronov2023holes,liu2023vertical}, usually requiring cryogenic temperatures. We use ab initio methods to obtain the electronic band structure and the Boltzmann transport equation in the relaxation time approximation to calculate the electron mobility. We anticipate that the electron mobility of Ge initially declines upon the introduction of Sn as a result of alloy scattering. Once the alloy transitions to a direct-gap material, however, we find that the mobility increases to many times that of intrinsic Ge.

\section{Methods}

The intrinsic electron mobility ($\mu_{\lambda\lambda}$) is given by solving the first iteration of the Boltzmann transport equation within the framework of the relaxation-time approximation \cite{Fisch}:

\begin{equation} \label{eq:mob}
\mu_{\lambda\lambda}^{(\alpha)} =\frac{2e}{3k_BTn^{(\alpha)}m_{\lambda}^{(\alpha)}}\int_{0}^{\infty} dE \ \frac{ E \rho^{(\alpha)}(E)}{R_{\lambda}^{(\alpha)}(E)} 
 \times f_0(E)(1+\gamma^{(\alpha)} E/2)   
\end{equation}

Here, $\rho^{(\alpha)}$ is the 3D density of states (DOS) per unit volume per spin at carrier energy $E$, $R_{\lambda}^{(\alpha)}$ is the total electron scattering rate, $n^{(\alpha)}$ is the carrier concentration and $m_{\lambda}^{(\alpha)}$ represents the effective masses. The superscript $(\alpha)$ indicates the conduction band valley, and the subscript $\lambda$ runs over the longitudinal and transverse directions since the effective masses are direction-dependent. At low temperatures, $f_0$ represents the Fermi-Dirac distribution, which simplifies to Boltzmann distribution at room temperature. The term $\gamma^\alpha$ is an isotropic non-parabolicity parameter, expressions for which can be found in \cite{fmathermo,jac,paige1964electrical}. Our calculated values correspond well with those calculated by Murphy-Armando for Ge \cite{fmathermo}. In our calculations, we only consider carriers populating the $L$- and $\Gamma$-valleys, since under our conditions the contribution from other valleys (i.e. $\Delta$ conduction valleys) is negligible.

The non-parabolic DOS per unit volume near the band minima are calculated using an ellipsoidal approximation:
\begin{widetext}
\begin{equation}
\rho^\alpha (E)=\frac{\sqrt{2}\sqrt{m_l^\alpha}{m_t^\alpha \sqrt{(E-E_c^\alpha)(1+\gamma^\alpha(E-E_c^\alpha))}}(1+2\gamma^\alpha(E-E_c^\alpha))}{\pi^2\hbar^3}\Theta(E-E_c^\alpha)
\label{eq:dos}
\end{equation}
\end{widetext}

where $m_l^\alpha$ ($m_t^\alpha$) are the longitudinal (transverse) effective masses respectively for valley $\alpha$, $\Theta(x)$ is the Heavyside function, where $\Theta(x)=1$ if $x>1$, $0$ otherwise, and $E_c^\alpha$ is the band edge of valley $\alpha$. The effective masses at $L$ were calculated for GeSn using the second derivative of the bandgap with respect to $k$ near $L$ in increments of $3\%$ concentration and fitted to a quadratic model. However, since the $\Gamma$ effective mass $m_\Gamma \rightarrow 0$ with increasing Sn concentration due to the increasing non-parabolicity of the bands around the $\Gamma$ point, equations \ref{eq:mob} and \ref{eq:dos} are no longer applicable for the $\Gamma$-mobility and $\Gamma$-DOS respectively. Therefore, we express the $\Gamma$-mobility as
\cite{ashcroft1976solid}:

\begin{equation} \label{eq:mobg}
\mu_{\lambda\lambda}^{(\alpha)} =\frac{e}{n^{(\alpha)}}\int_{0}^{\infty} \frac{d^3k}{4\pi^3} \ \frac{ \left(v_g^{(\alpha)}(k) \right)^2}{R_{\lambda}^{(\alpha)}\left(E(k)\right)} \left( \frac{-\partial f}{\partial E} \right),
\end{equation}

where $v_g^{(\alpha)}(k)\equiv \frac{1}{\hbar}\frac{\partial E}{\partial k}$ is the group velocity of the electrons at $\alpha$. By considering firstly that $\int d^3k=4\pi^3\int\rho(E)dE$,  and also for parabolic bands $\left(v_g^\alpha\right)^2=2E/m^\alpha$, it is easy to show that equation \ref{eq:mobg} is equivalent to equation \ref{eq:mob}. The DOS around the $\Gamma$-point was determined by $\rho^{(\alpha)}(E)\equiv \sum_k\delta\left( E-E_k \right)$. In order to calculate the group velocity and DOS around the $\Gamma$-point, we calculated the band energies $E_k$ around $\Gamma$ in GeSn using Density Functional Theory (DFT) and fitted the bands to a Kane model used to correct the band dispersion with the GW calculated band gap.

We calculated the bandgaps using the GW approximation \cite{Hedin1965,hybertsen1986electron}, which is implemented in ABINIT \cite{ab}. GW is a computationally heavy many-body perturbation theory method used to correct the underestimation in DFT band structure calculations. A deeper explanation is described by Aulbur \cite{GW}. Since GW calculations are performed at $0K$, we corrected the bandgap empirically for temperature using Varshni's empirical expression \cite{varshni1967temperature}. The effect of alloy content on the band edge is calculated using the Virtual Crystal Approximation (VCA), where the bandgap at each composition is taken to be $E_x^{VCA}=(1-x)E^{Ge}+xE^{Sn}+b^{GeSn}x(1-x)$. The bowing parameter $b$, which accounts for quadratic terms, is taken from Mukhopadhyay et al \cite{mukh}. Table \ref{tab:bandgaps} below displays the calculated bandgaps of Ge and Sn at room temperature.

\renewcommand{\arraystretch}{1.5}

\begin{table}[h!]
    \caption{Theoretically (th) calculated bandgaps (in eV) and effective masses (in $m_e$) of Ge and $\alpha$-Sn at room temperature ($300K$) along with experimental (exp) values.}
    \centering
    \begin{tabular}{ |c|c|c|c| }
     \hline 
    Parameter  & $Ge_{th}$   &  $Ge_{exp}$  &  $\alpha$-Sn \\  \hline
     $E_L$    &   0.66   & 0.66 \cite{precker2002experimental} & 0.09  \\
     $E_\Gamma$  &   0.8  &  0.8 \cite{el2010control}  & -0.41   \\
     $m^L_l$ & 1.636 & 1.64 \cite{dexter1956cyclotron}  & 1.83  \\
     $m^L_t$ & 0.074 & 0.08 \cite{dexter1956cyclotron} & 0.008  \\
     $m^\Gamma$ & 0.04 & 0.049 $\pm$ 0.07 \cite{clavelfma} & - \\
 \hline
    \end{tabular}
    \label{tab:bandgaps}
\end{table}

\subsection{Alloy scattering}

Alloy scattering and electron-phonon scattering make up the total scattering rate  $R_{\lambda}^{(\alpha)}(E)$ from equation \ref{eq:mob}. We calculate the phonon scattering by following Fischetti and Laux \cite{Fisch} using the deformation potential method. Since we are confining our calculations to low Sn compositions, we use the phonon energies and deformation potentials for Ge. The intra-valley parameters are taken from \cite{Fisch,fma08,hv}, and the inter-valley terms are taken from \cite{jac}. Elastic alloy scattering occurs due to the breaking of the translational symmetry by disorder. This effect can be studied from first-principles theory by introducing small changes in composition in a periodic crystal. The alloy scattering rate for an electron scattering from valley $\alpha$ to valley $\beta$ due to alloy disorder in the random substitutional alloy \cite{fma06} is given by:

\begin{equation}
R^\alpha(E)=\frac{2\pi}{\hbar}x(1-x)\frac{a_0^3}{8} \sum_{\beta} |\langle V_{\alpha\beta} \rangle|^2  \rho^{\beta}(E)
\end{equation}

where $x$ is the Sn concentration, $a_0$ is the lattice constant and $\langle V_{\alpha\beta} \rangle$ is the scattering matrix:

\begin{align} 
\left\langle V_{\alpha\beta} \right\rangle 
&= \left\langle V_{\alpha\beta}^{Sn} \right\rangle - \left\langle V_{\alpha\beta}^{Ge} \right\rangle \nonumber \\
&=N( \bra {\psi_\alpha} {\Delta V^{Sn}} \ket{\phi_{\beta}} - \bra{ \psi_\alpha }{ \Delta V^{Ge} } \ket{ \phi_{\beta} }) \label{pot}
\end{align}

In equation \ref{pot}, $N$ is the number of atoms in the supercell, $\psi$ represents the Bloch state of the periodic host lattice and $\phi$ is the exact eigenstate of the perturbing potential $\Delta V^A$, which is caused by the substitution of atom $A$ into the periodic host. The boundary condition $\psi(\vec{r})=\phi(\vec{r})$ is imposed when $\vec{r}$ is far from the type-A atom. $\Delta V^A$ tends to $0$ at a large distance from the impurity site. In a supercell calculation, the zero of the potential energy is arbitrary, so a physically well-defined method must be developed to compare the potentials of the supercell with $N$ host atoms and the supercell with one type-A atom and $N-1$ host atoms. We do this by calculating the difference between the average of the local DFT potential over points in the supercell far from the type-A atom and the same average in the periodic host. The difference in these averages gives the reference shift in the potentials and allows us to compare potentials and eigen-energies obtained in the two supercells \cite{fma08}. The supercell used to calculate the scattering rate must be large enough to allow for realistic structural relaxation of the host around the substitutional Ge or Sn atom. Also, the Bloch state wave vectors considered are limited by periodic boundary conditions being applied on the supercell single-particle wave functions. Supercells of sizes 64 and 128 atoms were used to find the scattering matrices between degenerate Bloch states.

The matrix elements $\left\langle V_{\alpha\beta} \right\rangle$ are calculated using first principles methods \cite{fma06}. We consider $17$ conduction band states, with only the intra-valley and inter-valley scattering parameters between $L$ and $\Gamma$ affecting our mobility calculation. Therefore, we seek four scattering matrix parameters: the intra-valley elements ($V_L$ and $V_\Gamma$),  the inter-valley term between the L valleys ($V_{LL}$) and the inter-valley term between $L$ and $\Gamma$ ($V_{L\Gamma}$ or $V_{L\Gamma}$). However, we do report on the intra-valley and g-type and f-type inter-valley scattering in the $X$ and $\frac{1}{2}X$ Brillouin zone points, since they become important to determine the scattering of the $\Delta$ conduction band valleys in measurements where an electric field causes electrons to accelerate to higher energies, or if GeSn is grown on Si \cite{harmin1982theory,paul20088}. $\left\langle V_{\alpha\beta} \right\rangle$ weakly depend on $x$ and can be considered as roughly independent of the host lattice at low Sn concentration \cite{fma06}. The lattice parameters for unstrained Ge and Sn are $5.66\AA$ and $6.49\AA$ respectively. The lattice parameters of unstrained and biaxially tensile strained GeSn were calculated at Sn concentrations up to $25\%$ in increments of $3\%$ Sn. The first-principles calculations are performed using Hartwigsen-Goedecker-Hutter \cite{hgh} plane-wave pseudopotentials in all of our Density Functional Theory calculations with ABINIT code \cite{ab}. We converge our results using an energy cut-off of 950eV and 8x8x8 Monkhorst-pack k-meshes for the first Brillouin zone integration. We apply the local density approximation for the exchange-correlation.

\section{Results}
\subsection{Band Structure}

\begin{figure}[h!]
    \includegraphics[width=0.8\textwidth]{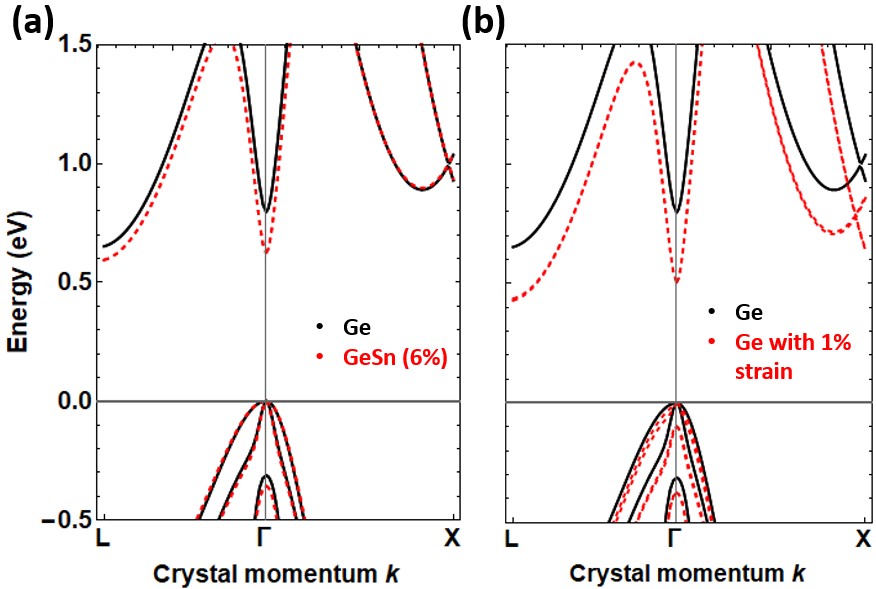} 
    \includegraphics[width=0.7\textwidth]{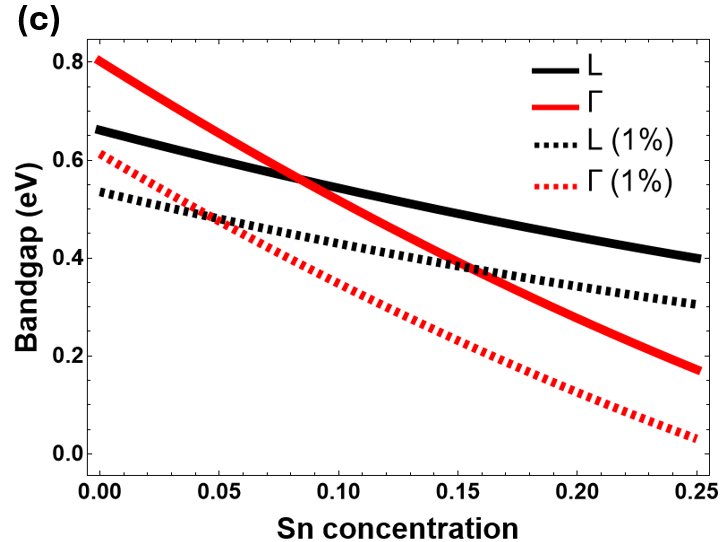}
    
    \caption{Electronic band structures of (a) Ge and $Ge_{0.94}Sn_{0.06}$, and (b) Ge and biaxially strained Ge at $1\%$ strain, as a function of crystallographic momentum $k$. (c): energy (eV) of $\Gamma$ (red) and $L$ (black) valleys in unstrained GeSn (solid) and $1\%$ tensile biaxial strained GeSn (dashed) against Sn concentration at $300K$.}
    \label{fig:crossover}
\end{figure}

As mentioned in the introduction, an increase in Sn concentration decreases the $\Gamma$-band energy at a faster rate than the indirect $L$-band energy, resulting in an indirect-to-direct bandgap transition. In  Figure \ref{fig:crossover}, we show the effect of (a) $6\%$ Sn and (b) $1\%$ biaxially strain on the bands of unstrained Ge. Figure \ref{fig:crossover}(c) shows the direct and indirect bandgaps of unstrained GeSn (solid) and $1\%$ biaxially strained GeSn (dashed) at $300K$. Figure \ref{fig:crossover}(c) shows that unstrained GeSn has direct bandgap above $8.4\%$ Sn concentration, in good agreement with the $6-9\%$ range determined in literature \cite{jiang,ghetmiri2014direct,gallagher2014compositional}. Sn concentrations of $7\%$ and $4.7\%$ are required to achieve a direct gap for $0.5\%$ and $1\%$ biaxially tensile strained GeSn, respectively. Figure \ref{fig:crossover} also shows that the gap closes for $1\%$ strain at around $25\%$ Sn. At such concentrations, the semiconductor would be in the strongly bi-polar regime. Since we are chiefly interested in the n-type transport in this work,  only report on mobilities at Sn concentrations up to $25\%$ in unstrained GeSn and $19\%$ in strained GeSn.


\subsection{Alloy scattering parameters}
Tables \ref{tab:intraVab} and \ref{tab:interVab} show the intra-valley and inter-valley alloy scattering parameters of GeSn respectively. We find that the inter- and intra-valley matrix elements are similar in magnitude. It is important to consider atomic relaxation near the impurity in the supercell, affecting the matrix elements by up to $1eV$ in magnitude. We observe that the alloy scattering parameters we calculate for GeSn are 3-4 times larger than those for SiGe \cite{fma06}, resulting in GeSn having alloy scattering 9-16 times stronger than SiGe.

\begin{table}[h!]
\caption{\label{tab:intraVab}Calculated intra-valley scattering matrix elements (in eV) for all relevant valleys in $Ge_{1-x}Sn_x$.} 
\centering
\begin{tabular}{ |c|c|c|c|c| } 
 \hline 
 Symbol  &  $V_L$       & $V_\Gamma$ &   $V_{\frac{1}{2}X}$   &    $V_X$ \\ 
 \hline
Value      &    -2.21      & -3.56            & -0.86                          &  -0.76\\ 
 \hline
\end{tabular}
\end{table}

\begin{table}[h!]
\caption{Calculated inter-valley scattering matrix elements (eV) for all relevant valleys in $Ge_{1-x}Sn_x$.} 
\centering
\begin{tabular}{ |c|c|c|c|c|c|c|c|c|c| } 
 \hline 
 Symbol  &  $V_{LL}$    &  $V_{L\Gamma}$ & $V_{LX}$ & $V_{\frac{1}{2}X_f}$  &  $V_{\frac{1}{2}X_g}$    &    $V_{X_f}$   &   $V_{X_g}$   &     $V_{\Gamma \frac{1}{2}X}$   &    $V_{\Gamma X}$ \\ 
 \hline
Value      &   -1.62     &     -2.93      &    0.83      &    -0.28      &      0.34        &     0.23      &     -0.47       &       -1.34        &   -1.73      \\ 
 \hline
\end{tabular}
\label{tab:interVab}
\end{table}

\subsection{Room temperature n-type electron mobility}

\begin{figure}[h!]
 \includegraphics[width=\linewidth]{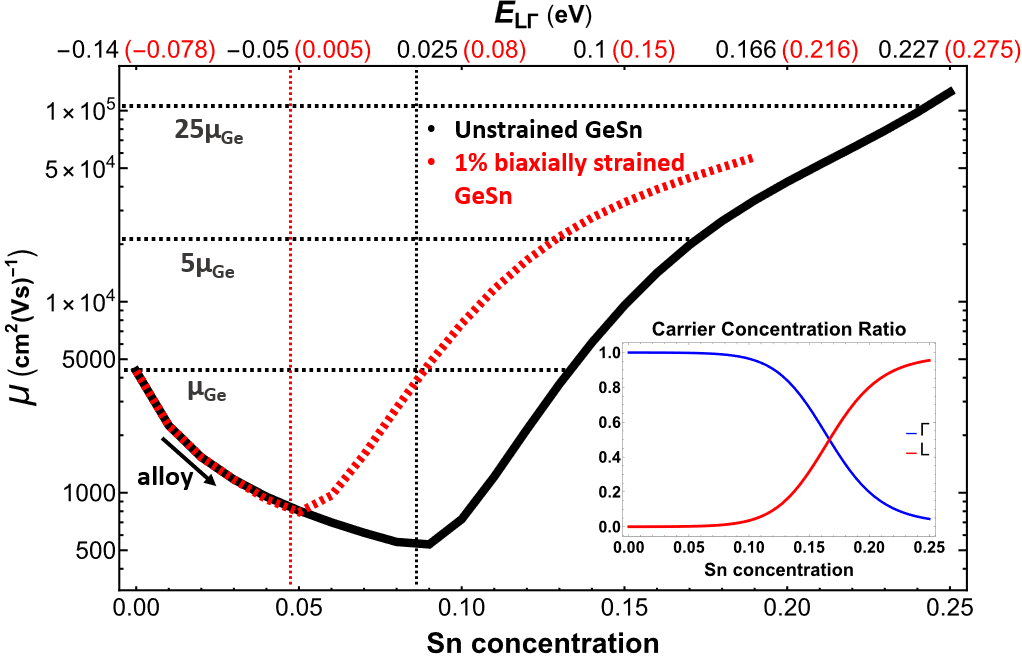}
    \caption{\label{fig:RTmob} Room-temperature intrinsic n-type electron mobility of unstrained GeSn (black) and $1\%$ biaxially tensile strained GeSn (red), as a function of Sn concentration.  The top axis shows $E_{L\Gamma}$, the energy difference (in eV) between $\Gamma$ and $L$. This scale is better representative of the behaviour of the electron mobility, since the mobility is strongly affected by $E_{L\Gamma}$, and experimentally Sn concentration may not correspond to $E_{L\Gamma}$ directly, for example due to Sn clustering.}
\end{figure}

Figure \ref{fig:RTmob} shows the calculated room temperature mobility of both unstrained GeSn (black) and $1\%$ biaxially strained GeSn (red) as a function of Sn concentration. Without strain, we obtain a mobility for Ge of $4300 \ cm^2/(Vs)$, which is comparable to the experimentally measured $3900 \ cm^2/(Vs)$\cite{Fisch}. As Sn is introduced, the mobility decreases due to alloy scattering, becoming as low as $510 \ cm^2/(Vs)$  at $8.75\%$ Sn (just beyond the indirect-to-direct gap transition, which we highlight with a vertical line). Thereafter, the mobility increases rapidly as the electrons begin to populate the $\Gamma$ valley, where the mobility is much higher compared to the $L$ valley. Figure \ref{fig:RTmob} also shows the carrier concentration ratio of the $\Gamma$ and $L$ valleys for intrinsic unstrained GeSn, which displays the gradual increase of electron population at $\Gamma$ due to the increasing energy difference with $L$. In order to achieve an electron mobility above that of Ge through alloying with Sn, a Sn concentration of at least $13.5\%$ must be added, resulting in $18\%$ carriers in the $\Gamma$ valley. Figure \ref{fig:RTmob} shows that at $17.5\%$ Sn, GeSn has a mobility $5$ times greater than that of Ge, while $Ge_{0.75}Sn_{0.25}$ has a mobility over $25$ times larger than for Ge. Despite this continuous increase in the mobility beyond $8.75\%$, it is nonetheless strongly limited by alloy scattering (see Appendix).

The behaviour of the mobility of GeSn under biaxial tensile strain is analogous to that of unstrained GeSn. The distinction is that strain induces an indirect-to-direct bandgap transition at lower Sn concentrations, resulting in less alloy scattering and therefore a higher electron mobility at lower Sn concentrations. In order to achieve a mobility higher than that of unstrained Ge, a Sn concentration of at least $11.8\%$ with $0.5\%$ strain is required, or $9\%$ Sn concentration with $1\%$ strain. In Figure \ref{fig:RTmob}, we see that under $1\%$ strain, a Sn concentration of just $13\%$ is required to achieve a mobility $5$ times higher than Ge. This could make biaxially strained GeSn more attractive than unstrained GeSn, given the segregation issues in GeSn.

\subsection{Low temperature (15K) n-type electron mobility}

\begin{figure}[h!]
    \includegraphics[width=\linewidth]{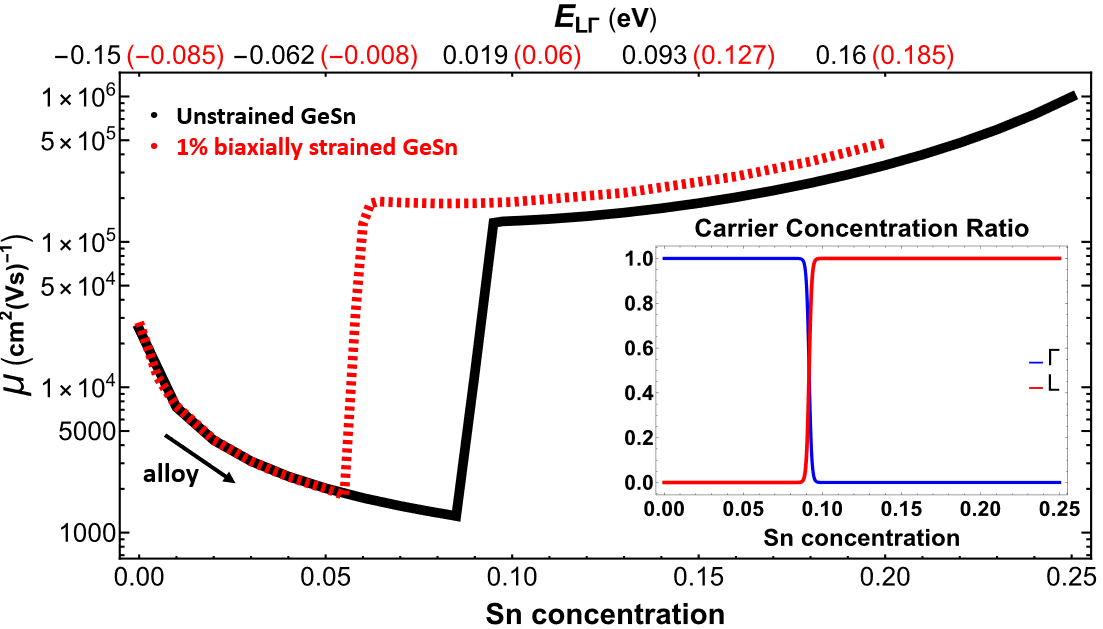}
    \caption{Low-temperature (15K) intrinsic n-type electron mobility and carrier concentration ratios at $L$ and $\Gamma$ as a function of Sn concentration and difference in bandgaps.}
    \label{fig:lowTmob}
\end{figure}

At $15K$, the electron distribution cannot be approximated by a Boltzmann distribution, and therefore $f_0(E)$ from Equation \ref{eq:mob} represents the Fermi-Dirac (FD) distribution at lower temperatures. Since at low temperatures the FD distribution behaves like a step-function, the mobility varies less smoothly compared to higher temperatures. At 15K, electron-phonon scattering is orders of magnitude lower than at room temperature. Unstrained, GeSn has a direct-gap beyond $8.8\%$ Sn concentration. Thereafter, electrons quickly populate the $\Gamma$ valley, causing the mobility to rise rapidly. In Figure \ref{fig:lowTmob}, we show the electron mobility of unstrained and $1\%$ biaxially strained GeSn, as well as the carrier concentration ratios of unstrained GeSn at $15K$. The electron mobility for unstrained Ge is $2.52\times 10^4 \ cm^2/(Vs)$, almost ten times the room temperature value. A Sn concentration of $9.1\%$ is needed to achieve a mobility higher than the Ge value, where the $L$ valley is $4.6 \ meV$ above the $\Gamma$ valley and the carrier concentration ratio of electrons in $\Gamma$ is $0.022$. The mobility reaches values in the order of $10^6 \ cm^2/(Vs)$ beyond $25\%$ Sn for unstrained GeSn. With the addition of tensile strain, a direct gap occurs at a lower Sn concentration, causing the mobility to increase at a lower Sn concentration, with less alloy scattering as a result. At $0.5\%$ and $1\%$ strain, GeSn transitions to a direct gap material at $7.4\%$ and $5.57\%$ Sn respectively. Beyond these concentrations, the mobility is almost immediately higher than for strained Ge. Under $1\%$ strain, as we see in Figure \ref{fig:lowTmob}, the mobility of GeSn is $10^5 \ cm^2/(Vs)$ at $5.95\%$ Sn, where $E_{L\Gamma} \simeq 0$ and the carrier concentration ratio of electrons in $\Gamma$ is $0.58$. At higher Sn concentrations, the effect of strain in increasing the mobility of GeSn in minimal.

\subsection{Effect of temperature on n-type electron mobility}

\begin{figure*}
    \includegraphics[width=\linewidth,keepaspectratio]{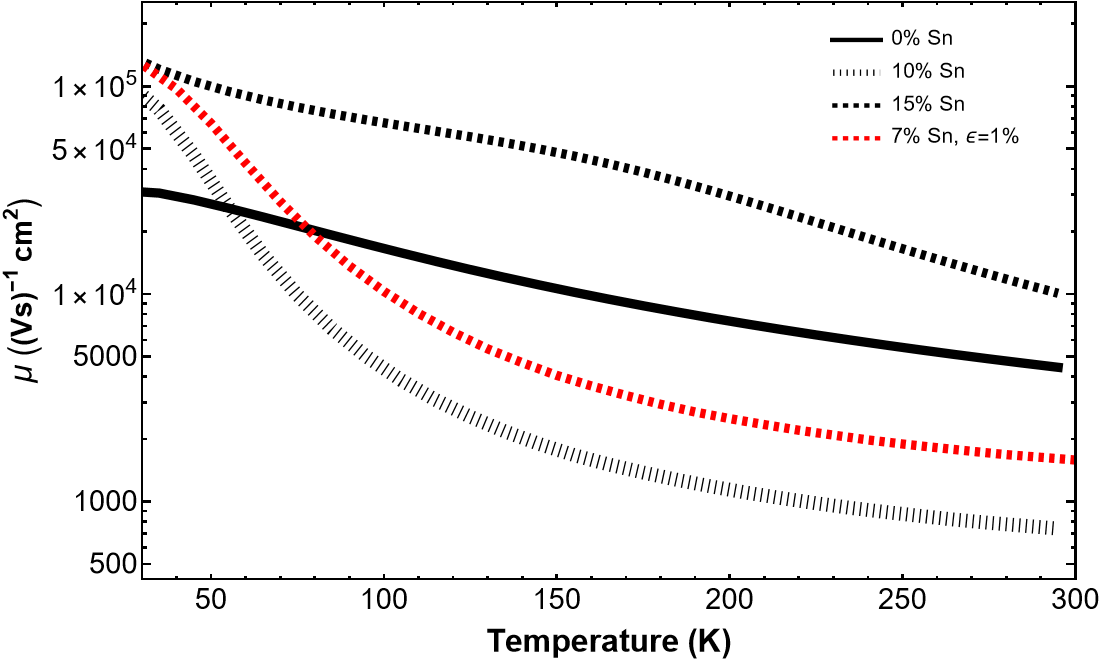}
   \caption{Intrinsic n-type electron mobility of Ge, $Ge_{0.9} Sn_{0.1}$ and $Ge_{0.85} Sn_{0.15}$, as well as $Ge_{0.93} Sn_{0.07}$ at $1\%$ biaxial tensile strain, as a function of temperature (K).}
    \label{fig:mobvT}
\end{figure*}

Reducing temperature increases the electron mobility of GeSn for two reasons. Firstly, electron-phonon scattering decreases with decreasing temperatures. Secondly,  the Fermi-distribution becomes increasingly step-function-like at lower temperatures, causing the carriers to populate the $\Gamma$ valley at the indirect-to-direct bandgap transition at lower $E_{L\Gamma}$ energy difference than at higher temperatures. This results in a higher electron mobility at lower Sn concentrations. From an experimental perspective it is easier to observe transport via the direct $\Gamma$ valley by varying the temperature, since this is easier to control than the Sn concentration. Therefore in Figure \ref{fig:mobvT}, we highlight the effect of temperature on the mobility of unstrained Ge, $Ge_{0.9}Sn_{0.1}$ and $Ge_{0.85}Sn_{0.15}$, as well as $Ge_{0.93}Sn_{0.07}$ under $1\%$ biaxial strain. Theoretically speaking, we can see from Figure \ref{fig:mobvT} that across the temperature range of $15-300K$, $Ge_{0.85}Sn_{0.15}$ provides the optimal mobility. However, as temperature decreases, the mobility of $Ge_{0.9}Sn_{0.1}$ becomes increasingly comparable to $Ge_{0.85}Sn_{0.15}$ and may be preferred experimentally to show the transition of $L\xrightarrow{} \Gamma$ transport. In addition, below $50K$, the mobility of $Ge_{0.93}Sn_{0.07}$ at $1\%$ biaxial strain is almost the same as unstrained $Ge_{0.9}Sn_{0.1}$, which may be even more desirable experimentally due to the lower Sn concentration.

\section{Conclusions}

In conclusion, we use first-principles methods to calculate the alloy scattering parameters of GeSn to determine the intrinsic n-type electron mobility of intrinsic GeSn as a function of alloy concentration and temperature. We find that alloy scattering plays a crucial role in the electron mobility, causing a severe drop in mobility to 10\% of that of Ge at low Sn concentrations up to 9\%.  However, once the indirect-to-direct bandgap transition occurs at higher Sn concentration, the mobility rises dramatically many times that of Ge. At room temperature, the electron mobility of unstrained GeSn can rise to above $25$ times higher than that of Ge at Sn concentrations above 25\%. The inclusion of biaxial tensile strain lowers the Sn content at which a high mobility can be achieved. At cryogenic temperatures, the dramatic increase in mobility occurs at the Sn concentration of the indirect-to-direct transition, requiring much less Sn than at room temperaure. We propose that measuring the mobility vs temperature in an alloy of 10\% Sn content will clearly show the dramatic effects of the indirect-to-direct transition. Given the promise of 2D hole gas mobilities in strained Ge, we believe strained GeSn holds promise as a material for the hosting of n-type 2D electron gases. The impact of these findings could lead to faster electronic devices with lower power consumption, particularly transistors and cryogenic electronics, and can guide the design of GeSn for optoelectronics.

\begin{acknowledgments}
This work was supported by Science Foundation Ireland: Grant No. 19/FFP/6953.
\end{acknowledgments}

\appendix

\section{Relative effect of alloy scattering on the mobility}

\begin{figure}[h!]
    \centering
    \includegraphics[width=\columnwidth]{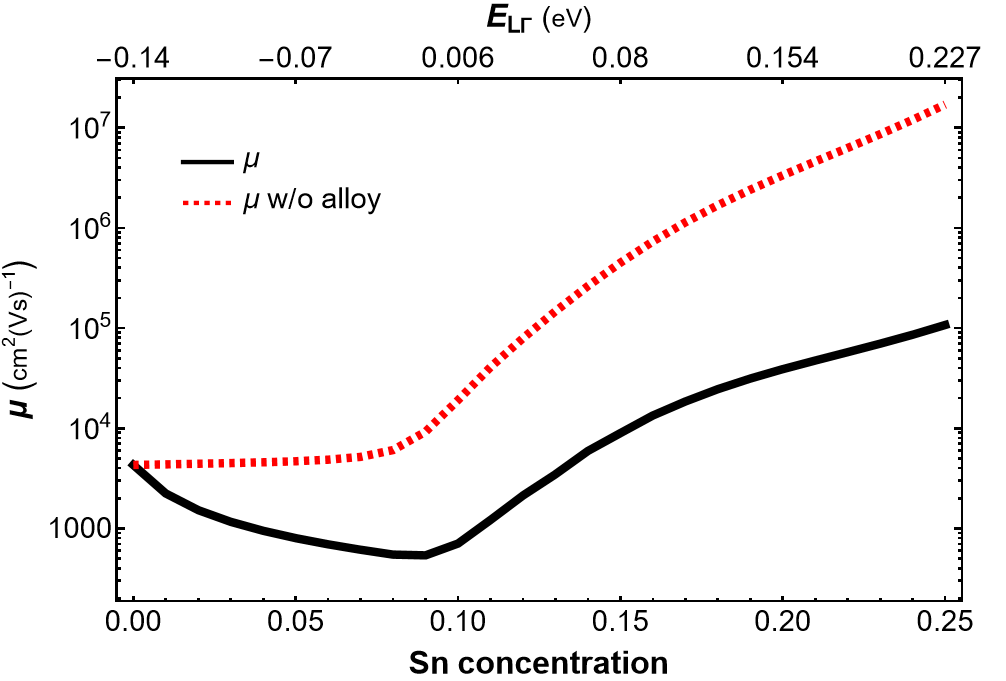}
    \caption{Room temperature mobility with and without alloy scattering contribution as a function of Sn concentration.}
    \label{fig:noalloy}
\end{figure}

Figure \ref{fig:noalloy} shows the relative effect of alloy scattering in the electron mobility. The black line corresponds to the mobility of GeSn at room temperature, as was shown in Figure \ref{fig:RTmob}. The red line is the hypothetical mobility disregarding alloy scattering. The fact that the outcomes are several orders of magnitude different highlights that both intra- and inter-valley alloy scattering need to be considered with precision rather than a single effective alloy scattering potential.

\begin{figure}[h!]
    \includegraphics[width=0.9\columnwidth]{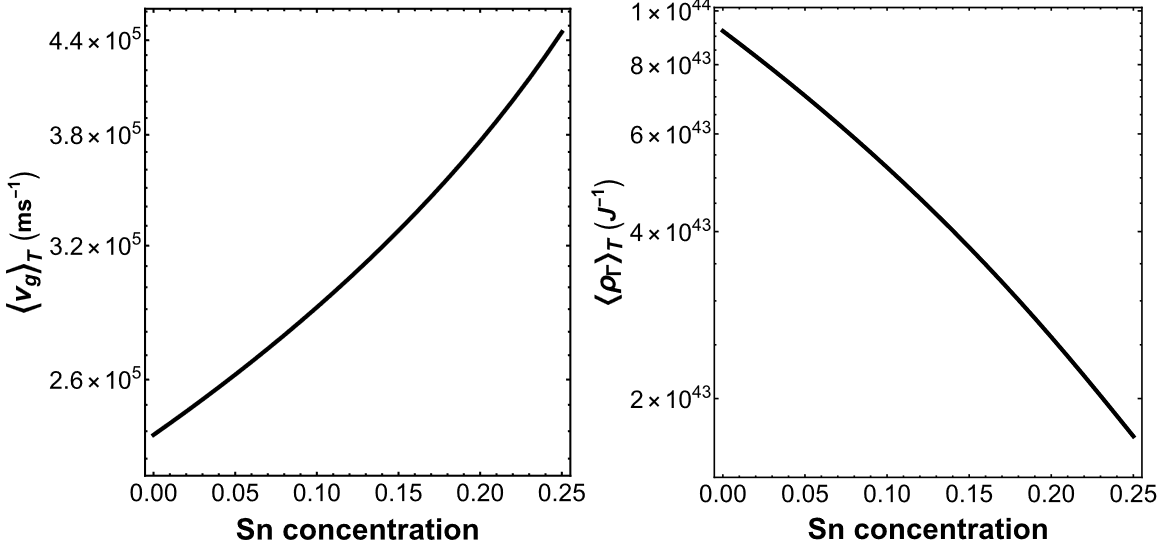}
    \caption{Thermal averages of the group velocity at $\Gamma$ in m/s (left) and the DOS (right) of GeSn as a function of Sn concentration.}
    \label{fig:thav}
\end{figure}

Another interesting feature of GeSn alloys, seen in Figures \ref{fig:RTmob}, \ref{fig:lowTmob} and \ref{fig:noalloy}, is that the mobility continues to rise when the carriers have almost entirely populated the $\Gamma$-valley, although one may expect and increasing effect of alloy scattering. This is due to two key reasons produced by increasing Sn concentration: (i)  an increase in the group velocities of the electrons, and (ii) the reduction of the DOS at $\Gamma$. In Figure \ref{fig:thav}, we show the thermal averages of the group velocity and $\Gamma$-DOS as functions of Sn concentration.

\begin{thebibliography}{65}%
\makeatletter
\providecommand \@ifxundefined [1]{%
 \@ifx{#1\undefined}
}%
\providecommand \@ifnum [1]{%
 \ifnum #1\expandafter \@firstoftwo
 \else \expandafter \@secondoftwo
 \fi
}%
\providecommand \@ifx [1]{%
 \ifx #1\expandafter \@firstoftwo
 \else \expandafter \@secondoftwo
 \fi
}%
\providecommand \natexlab [1]{#1}%
\providecommand \enquote  [1]{``#1''}%
\providecommand \bibnamefont  [1]{#1}%
\providecommand \bibfnamefont [1]{#1}%
\providecommand \citenamefont [1]{#1}%
\providecommand \href@noop [0]{\@secondoftwo}%
\providecommand \href [0]{\begingroup \@sanitize@url \@href}%
\providecommand \@href[1]{\@@startlink{#1}\@@href}%
\providecommand \@@href[1]{\endgroup#1\@@endlink}%
\providecommand \@sanitize@url [0]{\catcode `\\12\catcode `\$12\catcode `\&12\catcode `\#12\catcode `\^12\catcode `\_12\catcode `\%12\relax}%
\providecommand \@@startlink[1]{}%
\providecommand \@@endlink[0]{}%
\providecommand \url  [0]{\begingroup\@sanitize@url \@url }%
\providecommand \@url [1]{\endgroup\@href {#1}{\urlprefix }}%
\providecommand \urlprefix  [0]{URL }%
\providecommand \Eprint [0]{\href }%
\providecommand \doibase [0]{http://dx.doi.org/}%
\providecommand \selectlanguage [0]{\@gobble}%
\providecommand \bibinfo  [0]{\@secondoftwo}%
\providecommand \bibfield  [0]{\@secondoftwo}%
\providecommand \translation [1]{[#1]}%
\providecommand \BibitemOpen [0]{}%
\providecommand \bibitemStop [0]{}%
\providecommand \bibitemNoStop [0]{.\EOS\space}%
\providecommand \EOS [0]{\spacefactor3000\relax}%
\providecommand \BibitemShut  [1]{\csname bibitem#1\endcsname}%
\let\auto@bib@innerbib\@empty
\bibitem [{\citenamefont {Cardoux}\ \emph {et~al.}(2022)\citenamefont {Cardoux}, \citenamefont {Casiez}, \citenamefont {Pauc}, \citenamefont {Calvo}, \citenamefont {Coudurier}, \citenamefont {Rodriguez}, \citenamefont {Richy}, \citenamefont {Barritault}, \citenamefont {Lartigue}, \citenamefont {Constancias} \emph {et~al.}}]{cardoux2022room}%
  \BibitemOpen
  \bibfield  {author} {\bibinfo {author} {\bibfnamefont {C.}~\bibnamefont {Cardoux}}, \bibinfo {author} {\bibfnamefont {L.}~\bibnamefont {Casiez}}, \bibinfo {author} {\bibfnamefont {N.}~\bibnamefont {Pauc}}, \bibinfo {author} {\bibfnamefont {V.}~\bibnamefont {Calvo}}, \bibinfo {author} {\bibfnamefont {N.}~\bibnamefont {Coudurier}}, \bibinfo {author} {\bibfnamefont {P.}~\bibnamefont {Rodriguez}}, \bibinfo {author} {\bibfnamefont {J.}~\bibnamefont {Richy}}, \bibinfo {author} {\bibfnamefont {P.}~\bibnamefont {Barritault}}, \bibinfo {author} {\bibfnamefont {O.}~\bibnamefont {Lartigue}}, \bibinfo {author} {\bibfnamefont {C.}~\bibnamefont {Constancias}},  \emph {et~al.},\ }in\ \href@noop {} {\emph {\bibinfo {booktitle} {Silicon Photonics XVII}}},\ Vol.\ \bibinfo {volume} {12006}\ (\bibinfo {organization} {SPIE},\ \bibinfo {year} {2022})\ pp.\ \bibinfo {pages} {54--63}\BibitemShut {NoStop}%
\bibitem [{\citenamefont {Huang}\ \emph {et~al.}(2019)\citenamefont {Huang}, \citenamefont {Chang}, \citenamefont {Hsieh}, \citenamefont {Soref}, \citenamefont {Sun}, \citenamefont {Cheng},\ and\ \citenamefont {Chang}}]{huang2019electrically}%
  \BibitemOpen
  \bibfield  {author} {\bibinfo {author} {\bibfnamefont {B.-J.}\ \bibnamefont {Huang}}, \bibinfo {author} {\bibfnamefont {C.-Y.}\ \bibnamefont {Chang}}, \bibinfo {author} {\bibfnamefont {Y.-D.}\ \bibnamefont {Hsieh}}, \bibinfo {author} {\bibfnamefont {R.~A.}\ \bibnamefont {Soref}}, \bibinfo {author} {\bibfnamefont {G.}~\bibnamefont {Sun}}, \bibinfo {author} {\bibfnamefont {H.-H.}\ \bibnamefont {Cheng}}, \ and\ \bibinfo {author} {\bibfnamefont {G.-E.}\ \bibnamefont {Chang}},\ }\href@noop {} {\bibfield  {journal} {\bibinfo  {journal} {ACS Photonics}\ }\textbf {\bibinfo {volume} {6}},\ \bibinfo {pages} {1931} (\bibinfo {year} {2019})}\BibitemShut {NoStop}%
\bibitem [{\citenamefont {Chang}\ \emph {et~al.}(2017)\citenamefont {Chang}, \citenamefont {Chang}, \citenamefont {Li}, \citenamefont {Cheng}, \citenamefont {Soref}, \citenamefont {Sun},\ and\ \citenamefont {Hendrickson}}]{chang2017room}%
  \BibitemOpen
  \bibfield  {author} {\bibinfo {author} {\bibfnamefont {C.}~\bibnamefont {Chang}}, \bibinfo {author} {\bibfnamefont {T.-W.}\ \bibnamefont {Chang}}, \bibinfo {author} {\bibfnamefont {H.}~\bibnamefont {Li}}, \bibinfo {author} {\bibfnamefont {H.~H.}\ \bibnamefont {Cheng}}, \bibinfo {author} {\bibfnamefont {R.}~\bibnamefont {Soref}}, \bibinfo {author} {\bibfnamefont {G.}~\bibnamefont {Sun}}, \ and\ \bibinfo {author} {\bibfnamefont {J.~R.}\ \bibnamefont {Hendrickson}},\ }\href@noop {} {\bibfield  {journal} {\bibinfo  {journal} {Applied Physics Letters}\ }\textbf {\bibinfo {volume} {111}} (\bibinfo {year} {2017})}\BibitemShut {NoStop}%
\bibitem [{\citenamefont {Kim}\ \emph {et~al.}(2022)\citenamefont {Kim}, \citenamefont {Assali}, \citenamefont {Burt}, \citenamefont {Jung}, \citenamefont {Joo}, \citenamefont {Chen}, \citenamefont {Ikonic}, \citenamefont {Moutanabbir},\ and\ \citenamefont {Nam}}]{kim2022improved}%
  \BibitemOpen
  \bibfield  {author} {\bibinfo {author} {\bibfnamefont {Y.}~\bibnamefont {Kim}}, \bibinfo {author} {\bibfnamefont {S.}~\bibnamefont {Assali}}, \bibinfo {author} {\bibfnamefont {D.}~\bibnamefont {Burt}}, \bibinfo {author} {\bibfnamefont {Y.}~\bibnamefont {Jung}}, \bibinfo {author} {\bibfnamefont {H.-J.}\ \bibnamefont {Joo}}, \bibinfo {author} {\bibfnamefont {M.}~\bibnamefont {Chen}}, \bibinfo {author} {\bibfnamefont {Z.}~\bibnamefont {Ikonic}}, \bibinfo {author} {\bibfnamefont {O.}~\bibnamefont {Moutanabbir}}, \ and\ \bibinfo {author} {\bibfnamefont {D.}~\bibnamefont {Nam}},\ }in\ \href@noop {} {\emph {\bibinfo {booktitle} {Silicon Photonics XVII}}},\ Vol.\ \bibinfo {volume} {12006}\ (\bibinfo {organization} {SPIE},\ \bibinfo {year} {2022})\ pp.\ \bibinfo {pages} {151--157}\BibitemShut {NoStop}%
\bibitem [{\citenamefont {Ojo}\ \emph {et~al.}(2022)\citenamefont {Ojo}, \citenamefont {Zhou}, \citenamefont {Acharya}, \citenamefont {Saunders}, \citenamefont {Amoah}, \citenamefont {Jheng}, \citenamefont {Tran}, \citenamefont {Du}, \citenamefont {Chang}, \citenamefont {Li} \emph {et~al.}}]{ojo2022silicon}%
  \BibitemOpen
  \bibfield  {author} {\bibinfo {author} {\bibfnamefont {S.}~\bibnamefont {Ojo}}, \bibinfo {author} {\bibfnamefont {Y.}~\bibnamefont {Zhou}}, \bibinfo {author} {\bibfnamefont {S.}~\bibnamefont {Acharya}}, \bibinfo {author} {\bibfnamefont {N.}~\bibnamefont {Saunders}}, \bibinfo {author} {\bibfnamefont {S.}~\bibnamefont {Amoah}}, \bibinfo {author} {\bibfnamefont {Y.-T.}\ \bibnamefont {Jheng}}, \bibinfo {author} {\bibfnamefont {H.}~\bibnamefont {Tran}}, \bibinfo {author} {\bibfnamefont {W.}~\bibnamefont {Du}}, \bibinfo {author} {\bibfnamefont {G.-E.}\ \bibnamefont {Chang}}, \bibinfo {author} {\bibfnamefont {B.}~\bibnamefont {Li}},  \emph {et~al.},\ }in\ \href@noop {} {\emph {\bibinfo {booktitle} {Physics and Simulation of Optoelectronic Devices XXX}}},\ Vol.\ \bibinfo {volume} {11995}\ (\bibinfo {organization} {SPIE},\ \bibinfo {year} {2022})\ pp.\ \bibinfo {pages} {78--83}\BibitemShut {NoStop}%
\bibitem [{\citenamefont {Marzban}\ \emph {et~al.}(2022)\citenamefont {Marzban}, \citenamefont {Seidel}, \citenamefont {Liu}, \citenamefont {Wu}, \citenamefont {Kiyek}, \citenamefont {Zoellner}, \citenamefont {Ikonic}, \citenamefont {Schulze}, \citenamefont {GruÌˆtzmacher}, \citenamefont {Capellini} \emph {et~al.}}]{marzban2022strain}%
  \BibitemOpen
  \bibfield  {author} {\bibinfo {author} {\bibfnamefont {B.}~\bibnamefont {Marzban}}, \bibinfo {author} {\bibfnamefont {L.}~\bibnamefont {Seidel}}, \bibinfo {author} {\bibfnamefont {T.}~\bibnamefont {Liu}}, \bibinfo {author} {\bibfnamefont {K.}~\bibnamefont {Wu}}, \bibinfo {author} {\bibfnamefont {V.}~\bibnamefont {Kiyek}}, \bibinfo {author} {\bibfnamefont {M.~H.}\ \bibnamefont {Zoellner}}, \bibinfo {author} {\bibfnamefont {Z.}~\bibnamefont {Ikonic}}, \bibinfo {author} {\bibfnamefont {J.}~\bibnamefont {Schulze}}, \bibinfo {author} {\bibfnamefont {D.}~\bibnamefont {GruÌˆtzmacher}}, \bibinfo {author} {\bibfnamefont {G.}~\bibnamefont {Capellini}},  \emph {et~al.},\ }\href@noop {} {\bibfield  {journal} {\bibinfo  {journal} {ACS Photonics}\ }\textbf {\bibinfo {volume} {10}},\ \bibinfo {pages} {217} (\bibinfo {year} {2022})}\BibitemShut {NoStop}%
\bibitem [{\citenamefont {Wirths}\ \emph {et~al.}(2015)\citenamefont {Wirths}, \citenamefont {Geiger}, \citenamefont {Von Den~Driesch}, \citenamefont {Mussler}, \citenamefont {Stoica}, \citenamefont {Mantl}, \citenamefont {Ikonic}, \citenamefont {Luysberg}, \citenamefont {Chiussi}, \citenamefont {Hartmann} \emph {et~al.}}]{wirths2015lasing}%
  \BibitemOpen
  \bibfield  {author} {\bibinfo {author} {\bibfnamefont {S.}~\bibnamefont {Wirths}}, \bibinfo {author} {\bibfnamefont {R.}~\bibnamefont {Geiger}}, \bibinfo {author} {\bibfnamefont {N.}~\bibnamefont {Von Den~Driesch}}, \bibinfo {author} {\bibfnamefont {G.}~\bibnamefont {Mussler}}, \bibinfo {author} {\bibfnamefont {T.}~\bibnamefont {Stoica}}, \bibinfo {author} {\bibfnamefont {S.}~\bibnamefont {Mantl}}, \bibinfo {author} {\bibfnamefont {Z.}~\bibnamefont {Ikonic}}, \bibinfo {author} {\bibfnamefont {M.}~\bibnamefont {Luysberg}}, \bibinfo {author} {\bibfnamefont {S.}~\bibnamefont {Chiussi}}, \bibinfo {author} {\bibfnamefont {J.-M.}\ \bibnamefont {Hartmann}},  \emph {et~al.},\ }\href@noop {} {\bibfield  {journal} {\bibinfo  {journal} {Nature photonics}\ }\textbf {\bibinfo {volume} {9}},\ \bibinfo {pages} {88} (\bibinfo {year} {2015})}\BibitemShut {NoStop}%
\bibitem [{\citenamefont {Atalla}\ \emph {et~al.}(2022)\citenamefont {Atalla}, \citenamefont {Assali}, \citenamefont {Koelling}, \citenamefont {Attiaoui},\ and\ \citenamefont {Moutanabbir}}]{atalla2022high}%
  \BibitemOpen
  \bibfield  {author} {\bibinfo {author} {\bibfnamefont {M.~R.}\ \bibnamefont {Atalla}}, \bibinfo {author} {\bibfnamefont {S.}~\bibnamefont {Assali}}, \bibinfo {author} {\bibfnamefont {S.}~\bibnamefont {Koelling}}, \bibinfo {author} {\bibfnamefont {A.}~\bibnamefont {Attiaoui}}, \ and\ \bibinfo {author} {\bibfnamefont {O.}~\bibnamefont {Moutanabbir}},\ }\href@noop {} {\bibfield  {journal} {\bibinfo  {journal} {ACS Photonics}\ }\textbf {\bibinfo {volume} {9}},\ \bibinfo {pages} {1425} (\bibinfo {year} {2022})}\BibitemShut {NoStop}%
\bibitem [{\citenamefont {Talamas~Simola}\ \emph {et~al.}(2021)\citenamefont {Talamas~Simola}, \citenamefont {Kiyek}, \citenamefont {Ballabio}, \citenamefont {Schlykow}, \citenamefont {Frigerio}, \citenamefont {Zucchetti}, \citenamefont {De~Iacovo}, \citenamefont {Colace}, \citenamefont {Yamamoto}, \citenamefont {Capellini} \emph {et~al.}}]{talamas2021cmos}%
  \BibitemOpen
  \bibfield  {author} {\bibinfo {author} {\bibfnamefont {E.}~\bibnamefont {Talamas~Simola}}, \bibinfo {author} {\bibfnamefont {V.}~\bibnamefont {Kiyek}}, \bibinfo {author} {\bibfnamefont {A.}~\bibnamefont {Ballabio}}, \bibinfo {author} {\bibfnamefont {V.}~\bibnamefont {Schlykow}}, \bibinfo {author} {\bibfnamefont {J.}~\bibnamefont {Frigerio}}, \bibinfo {author} {\bibfnamefont {C.}~\bibnamefont {Zucchetti}}, \bibinfo {author} {\bibfnamefont {A.}~\bibnamefont {De~Iacovo}}, \bibinfo {author} {\bibfnamefont {L.}~\bibnamefont {Colace}}, \bibinfo {author} {\bibfnamefont {Y.}~\bibnamefont {Yamamoto}}, \bibinfo {author} {\bibfnamefont {G.}~\bibnamefont {Capellini}},  \emph {et~al.},\ }\href@noop {} {\bibfield  {journal} {\bibinfo  {journal} {ACS photonics}\ }\textbf {\bibinfo {volume} {8}},\ \bibinfo {pages} {2166} (\bibinfo {year} {2021})}\BibitemShut {NoStop}%
\bibitem [{\citenamefont {Tran}\ \emph {et~al.}(2019)\citenamefont {Tran}, \citenamefont {Pham}, \citenamefont {Margetis}, \citenamefont {Zhou}, \citenamefont {Dou}, \citenamefont {Grant}, \citenamefont {Grant}, \citenamefont {Al-Kabi}, \citenamefont {Sun}, \citenamefont {Soref} \emph {et~al.}}]{tran2019si}%
  \BibitemOpen
  \bibfield  {author} {\bibinfo {author} {\bibfnamefont {H.}~\bibnamefont {Tran}}, \bibinfo {author} {\bibfnamefont {T.}~\bibnamefont {Pham}}, \bibinfo {author} {\bibfnamefont {J.}~\bibnamefont {Margetis}}, \bibinfo {author} {\bibfnamefont {Y.}~\bibnamefont {Zhou}}, \bibinfo {author} {\bibfnamefont {W.}~\bibnamefont {Dou}}, \bibinfo {author} {\bibfnamefont {P.~C.}\ \bibnamefont {Grant}}, \bibinfo {author} {\bibfnamefont {J.~M.}\ \bibnamefont {Grant}}, \bibinfo {author} {\bibfnamefont {S.}~\bibnamefont {Al-Kabi}}, \bibinfo {author} {\bibfnamefont {G.}~\bibnamefont {Sun}}, \bibinfo {author} {\bibfnamefont {R.~A.}\ \bibnamefont {Soref}},  \emph {et~al.},\ }\href@noop {} {\bibfield  {journal} {\bibinfo  {journal} {ACS Photonics}\ }\textbf {\bibinfo {volume} {6}},\ \bibinfo {pages} {2807} (\bibinfo {year} {2019})}\BibitemShut {NoStop}%
\bibitem [{\citenamefont {Zhang}\ \emph {et~al.}(2018)\citenamefont {Zhang}, \citenamefont {Hu}, \citenamefont {Liu}, \citenamefont {Lin}, \citenamefont {Wang}, \citenamefont {Ding}, \citenamefont {Wang}, \citenamefont {Cheng},\ and\ \citenamefont {Xue}}]{zhang2018gesn}%
  \BibitemOpen
  \bibfield  {author} {\bibinfo {author} {\bibfnamefont {D.}~\bibnamefont {Zhang}}, \bibinfo {author} {\bibfnamefont {X.}~\bibnamefont {Hu}}, \bibinfo {author} {\bibfnamefont {D.}~\bibnamefont {Liu}}, \bibinfo {author} {\bibfnamefont {X.}~\bibnamefont {Lin}}, \bibinfo {author} {\bibfnamefont {W.}~\bibnamefont {Wang}}, \bibinfo {author} {\bibfnamefont {Z.}~\bibnamefont {Ding}}, \bibinfo {author} {\bibfnamefont {Z.}~\bibnamefont {Wang}}, \bibinfo {author} {\bibfnamefont {B.}~\bibnamefont {Cheng}}, \ and\ \bibinfo {author} {\bibfnamefont {C.}~\bibnamefont {Xue}},\ }in\ \href@noop {} {\emph {\bibinfo {booktitle} {Optical Sensing and Imaging Technologies and Applications}}},\ Vol.\ \bibinfo {volume} {10846}\ (\bibinfo {organization} {SPIE},\ \bibinfo {year} {2018})\ pp.\ \bibinfo {pages} {327--334}\BibitemShut {NoStop}%
\bibitem [{\citenamefont {Miller}(2009)}]{miller2009device}%
  \BibitemOpen
  \bibfield  {author} {\bibinfo {author} {\bibfnamefont {D.~A.}\ \bibnamefont {Miller}},\ }\href@noop {} {\bibfield  {journal} {\bibinfo  {journal} {Proceedings of the IEEE}\ }\textbf {\bibinfo {volume} {97}},\ \bibinfo {pages} {1166} (\bibinfo {year} {2009})}\BibitemShut {NoStop}%
\bibitem [{\citenamefont {Liu}\ \emph {et~al.}(2018)\citenamefont {Liu}, \citenamefont {Deng}, \citenamefont {Zhang},\ and\ \citenamefont {Yao}}]{liu2018chip}%
  \BibitemOpen
  \bibfield  {author} {\bibinfo {author} {\bibfnamefont {J.}~\bibnamefont {Liu}}, \bibinfo {author} {\bibfnamefont {H.}~\bibnamefont {Deng}}, \bibinfo {author} {\bibfnamefont {W.}~\bibnamefont {Zhang}}, \ and\ \bibinfo {author} {\bibfnamefont {J.}~\bibnamefont {Yao}},\ }\href@noop {} {\bibfield  {journal} {\bibinfo  {journal} {Journal of Lightwave Technology}\ }\textbf {\bibinfo {volume} {36}},\ \bibinfo {pages} {4099} (\bibinfo {year} {2018})}\BibitemShut {NoStop}%
\bibitem [{\citenamefont {Lavchiev}\ and\ \citenamefont {Jakoby}(2016)}]{lavchiev2016photonics}%
  \BibitemOpen
  \bibfield  {author} {\bibinfo {author} {\bibfnamefont {V.~M.}\ \bibnamefont {Lavchiev}}\ and\ \bibinfo {author} {\bibfnamefont {B.}~\bibnamefont {Jakoby}},\ }\href@noop {} {\bibfield  {journal} {\bibinfo  {journal} {IEEE Journal of Selected Topics in Quantum Electronics}\ }\textbf {\bibinfo {volume} {23}},\ \bibinfo {pages} {452} (\bibinfo {year} {2016})}\BibitemShut {NoStop}%
\bibitem [{\citenamefont {Dutt}\ \emph {et~al.}(2013)\citenamefont {Dutt}, \citenamefont {Lin}, \citenamefont {Sukhdeo}, \citenamefont {Vulovic}, \citenamefont {Gupta}, \citenamefont {Nam}, \citenamefont {Saraswat},\ and\ \citenamefont {Harris~Jr}}]{dutt}%
  \BibitemOpen
  \bibfield  {author} {\bibinfo {author} {\bibfnamefont {B.}~\bibnamefont {Dutt}}, \bibinfo {author} {\bibfnamefont {H.}~\bibnamefont {Lin}}, \bibinfo {author} {\bibfnamefont {D.~S.}\ \bibnamefont {Sukhdeo}}, \bibinfo {author} {\bibfnamefont {B.~M.}\ \bibnamefont {Vulovic}}, \bibinfo {author} {\bibfnamefont {S.}~\bibnamefont {Gupta}}, \bibinfo {author} {\bibfnamefont {D.}~\bibnamefont {Nam}}, \bibinfo {author} {\bibfnamefont {K.~C.}\ \bibnamefont {Saraswat}}, \ and\ \bibinfo {author} {\bibfnamefont {J.~S.}\ \bibnamefont {Harris~Jr}},\ }\href@noop {} {\bibfield  {journal} {\bibinfo  {journal} {IEEE Journal of Selected Topics in Quantum Electronics}\ }\textbf {\bibinfo {volume} {19}},\ \bibinfo {pages} {1502706} (\bibinfo {year} {2013})}\BibitemShut {NoStop}%
\bibitem [{\citenamefont {Liu}\ \emph {et~al.}(2007)\citenamefont {Liu}, \citenamefont {Sun}, \citenamefont {Pan}, \citenamefont {Wang}, \citenamefont {Kimerling}, \citenamefont {Koch},\ and\ \citenamefont {Michel}}]{liu2007tensile}%
  \BibitemOpen
  \bibfield  {author} {\bibinfo {author} {\bibfnamefont {J.}~\bibnamefont {Liu}}, \bibinfo {author} {\bibfnamefont {X.}~\bibnamefont {Sun}}, \bibinfo {author} {\bibfnamefont {D.}~\bibnamefont {Pan}}, \bibinfo {author} {\bibfnamefont {X.}~\bibnamefont {Wang}}, \bibinfo {author} {\bibfnamefont {L.~C.}\ \bibnamefont {Kimerling}}, \bibinfo {author} {\bibfnamefont {T.~L.}\ \bibnamefont {Koch}}, \ and\ \bibinfo {author} {\bibfnamefont {J.}~\bibnamefont {Michel}},\ }\href@noop {} {\bibfield  {journal} {\bibinfo  {journal} {Optics express}\ }\textbf {\bibinfo {volume} {15}},\ \bibinfo {pages} {11272} (\bibinfo {year} {2007})}\BibitemShut {NoStop}%
\bibitem [{\citenamefont {Wirths}\ \emph {et~al.}(2013)\citenamefont {Wirths}, \citenamefont {Ikonic}, \citenamefont {Tiedemann}, \citenamefont {Holl{\"a}nder}, \citenamefont {Stoica}, \citenamefont {Mussler}, \citenamefont {Breuer}, \citenamefont {Hartmann}, \citenamefont {Benedetti}, \citenamefont {Chiussi} \emph {et~al.}}]{wirths2013tensely}%
  \BibitemOpen
  \bibfield  {author} {\bibinfo {author} {\bibfnamefont {S.}~\bibnamefont {Wirths}}, \bibinfo {author} {\bibfnamefont {Z.}~\bibnamefont {Ikonic}}, \bibinfo {author} {\bibfnamefont {A.}~\bibnamefont {Tiedemann}}, \bibinfo {author} {\bibfnamefont {B.}~\bibnamefont {Holl{\"a}nder}}, \bibinfo {author} {\bibfnamefont {T.}~\bibnamefont {Stoica}}, \bibinfo {author} {\bibfnamefont {G.}~\bibnamefont {Mussler}}, \bibinfo {author} {\bibfnamefont {U.}~\bibnamefont {Breuer}}, \bibinfo {author} {\bibfnamefont {J.}~\bibnamefont {Hartmann}}, \bibinfo {author} {\bibfnamefont {A.}~\bibnamefont {Benedetti}}, \bibinfo {author} {\bibfnamefont {S.}~\bibnamefont {Chiussi}},  \emph {et~al.},\ }\href@noop {} {\bibfield  {journal} {\bibinfo  {journal} {Applied physics letters}\ }\textbf {\bibinfo {volume} {103}} (\bibinfo {year} {2013})}\BibitemShut {NoStop}%
\bibitem [{\citenamefont {Giunto}\ \emph {et~al.}(2023)\citenamefont {Giunto} \emph {et~al.}}]{giunto2023gesn}%
  \BibitemOpen
  \bibfield  {author} {\bibinfo {author} {\bibfnamefont {A.}~\bibnamefont {Giunto}} \emph {et~al.},\ }\href@noop {} {\bibfield  {journal} {\bibinfo  {journal} {arXiv preprint arXiv:2309.10584}\ } (\bibinfo {year} {2023})}\BibitemShut {NoStop}%
\bibitem [{\citenamefont {Moutanabbir}\ \emph {et~al.}(2021)\citenamefont {Moutanabbir}, \citenamefont {Assali}, \citenamefont {Gong}, \citenamefont {O'Reilly}, \citenamefont {Broderick}, \citenamefont {Marzban}, \citenamefont {Witzens}, \citenamefont {Du}, \citenamefont {Yu}, \citenamefont {Chelnokov} \emph {et~al.}}]{moutanabbir2021monolithic}%
  \BibitemOpen
  \bibfield  {author} {\bibinfo {author} {\bibfnamefont {O.}~\bibnamefont {Moutanabbir}}, \bibinfo {author} {\bibfnamefont {S.}~\bibnamefont {Assali}}, \bibinfo {author} {\bibfnamefont {X.}~\bibnamefont {Gong}}, \bibinfo {author} {\bibfnamefont {E.}~\bibnamefont {O'Reilly}}, \bibinfo {author} {\bibfnamefont {C.}~\bibnamefont {Broderick}}, \bibinfo {author} {\bibfnamefont {B.}~\bibnamefont {Marzban}}, \bibinfo {author} {\bibfnamefont {J.}~\bibnamefont {Witzens}}, \bibinfo {author} {\bibfnamefont {W.}~\bibnamefont {Du}}, \bibinfo {author} {\bibfnamefont {S.-Q.}\ \bibnamefont {Yu}}, \bibinfo {author} {\bibfnamefont {A.}~\bibnamefont {Chelnokov}},  \emph {et~al.},\ }\href@noop {} {\bibfield  {journal} {\bibinfo  {journal} {Applied Physics Letters}\ }\textbf {\bibinfo {volume} {118}} (\bibinfo {year} {2021})}\BibitemShut {NoStop}%
\bibitem [{\citenamefont {Sau}\ and\ \citenamefont {Cohen}(2007)}]{sau}%
  \BibitemOpen
  \bibfield  {author} {\bibinfo {author} {\bibfnamefont {J.~D.}\ \bibnamefont {Sau}}\ and\ \bibinfo {author} {\bibfnamefont {M.~L.}\ \bibnamefont {Cohen}},\ }\href@noop {} {\bibfield  {journal} {\bibinfo  {journal} {Physical Review B}\ }\textbf {\bibinfo {volume} {75}},\ \bibinfo {pages} {045208} (\bibinfo {year} {2007})}\BibitemShut {NoStop}%
\bibitem [{\citenamefont {Mukhopadhyay}\ \emph {et~al.}(2017)\citenamefont {Mukhopadhyay}, \citenamefont {Sen}, \citenamefont {Basu}, \citenamefont {Mukhopadhyay},\ and\ \citenamefont {Basu}}]{mukh}%
  \BibitemOpen
  \bibfield  {author} {\bibinfo {author} {\bibfnamefont {B.}~\bibnamefont {Mukhopadhyay}}, \bibinfo {author} {\bibfnamefont {G.}~\bibnamefont {Sen}}, \bibinfo {author} {\bibfnamefont {R.}~\bibnamefont {Basu}}, \bibinfo {author} {\bibfnamefont {S.}~\bibnamefont {Mukhopadhyay}}, \ and\ \bibinfo {author} {\bibfnamefont {P.~K.}\ \bibnamefont {Basu}},\ }\href@noop {} {\bibfield  {journal} {\bibinfo  {journal} {physica status solidi (b)}\ }\textbf {\bibinfo {volume} {254}},\ \bibinfo {pages} {1700244} (\bibinfo {year} {2017})}\BibitemShut {NoStop}%
\bibitem [{\citenamefont {Chuang}\ \emph {et~al.}(2020)\citenamefont {Chuang}, \citenamefont {Liu}, \citenamefont {Luo},\ and\ \citenamefont {Li}}]{chuang2020electron}%
  \BibitemOpen
  \bibfield  {author} {\bibinfo {author} {\bibfnamefont {Y.}~\bibnamefont {Chuang}}, \bibinfo {author} {\bibfnamefont {C.-Y.}\ \bibnamefont {Liu}}, \bibinfo {author} {\bibfnamefont {G.-L.}\ \bibnamefont {Luo}}, \ and\ \bibinfo {author} {\bibfnamefont {J.-Y.}\ \bibnamefont {Li}},\ }\href@noop {} {\bibfield  {journal} {\bibinfo  {journal} {IEEE Electron Device Letters}\ }\textbf {\bibinfo {volume} {42}},\ \bibinfo {pages} {10} (\bibinfo {year} {2020})}\BibitemShut {NoStop}%
\bibitem [{\citenamefont {Huang}\ \emph {et~al.}(2017)\citenamefont {Huang}, \citenamefont {Tsou}, \citenamefont {Huang}, \citenamefont {Huang}, \citenamefont {Lan}, \citenamefont {Liu}, \citenamefont {Huang}, \citenamefont {Chung}, \citenamefont {Chang}, \citenamefont {Chu} \emph {et~al.}}]{huang2017high}%
  \BibitemOpen
  \bibfield  {author} {\bibinfo {author} {\bibfnamefont {Y.-S.}\ \bibnamefont {Huang}}, \bibinfo {author} {\bibfnamefont {Y.-J.}\ \bibnamefont {Tsou}}, \bibinfo {author} {\bibfnamefont {C.-H.}\ \bibnamefont {Huang}}, \bibinfo {author} {\bibfnamefont {C.-H.}\ \bibnamefont {Huang}}, \bibinfo {author} {\bibfnamefont {H.-S.}\ \bibnamefont {Lan}}, \bibinfo {author} {\bibfnamefont {C.~W.}\ \bibnamefont {Liu}}, \bibinfo {author} {\bibfnamefont {Y.-C.}\ \bibnamefont {Huang}}, \bibinfo {author} {\bibfnamefont {H.}~\bibnamefont {Chung}}, \bibinfo {author} {\bibfnamefont {C.-P.}\ \bibnamefont {Chang}}, \bibinfo {author} {\bibfnamefont {S.~S.}\ \bibnamefont {Chu}},  \emph {et~al.},\ }\href@noop {} {\bibfield  {journal} {\bibinfo  {journal} {IEEE Transactions on Electron Devices}\ }\textbf {\bibinfo {volume} {64}},\ \bibinfo {pages} {2498} (\bibinfo {year} {2017})}\BibitemShut {NoStop}%
\bibitem [{\citenamefont {Scappucci}\ \emph {et~al.}(2021)\citenamefont {Scappucci}, \citenamefont {Kloeffel}, \citenamefont {Zwanenburg}, \citenamefont {Loss}, \citenamefont {Myronov}, \citenamefont {Zhang}, \citenamefont {De~Franceschi}, \citenamefont {Katsaros},\ and\ \citenamefont {Veldhorst}}]{scappucci2021germanium}%
  \BibitemOpen
  \bibfield  {author} {\bibinfo {author} {\bibfnamefont {G.}~\bibnamefont {Scappucci}}, \bibinfo {author} {\bibfnamefont {C.}~\bibnamefont {Kloeffel}}, \bibinfo {author} {\bibfnamefont {F.~A.}\ \bibnamefont {Zwanenburg}}, \bibinfo {author} {\bibfnamefont {D.}~\bibnamefont {Loss}}, \bibinfo {author} {\bibfnamefont {M.}~\bibnamefont {Myronov}}, \bibinfo {author} {\bibfnamefont {J.-J.}\ \bibnamefont {Zhang}}, \bibinfo {author} {\bibfnamefont {S.}~\bibnamefont {De~Franceschi}}, \bibinfo {author} {\bibfnamefont {G.}~\bibnamefont {Katsaros}}, \ and\ \bibinfo {author} {\bibfnamefont {M.}~\bibnamefont {Veldhorst}},\ }\href@noop {} {\bibfield  {journal} {\bibinfo  {journal} {Nature Reviews Materials}\ }\textbf {\bibinfo {volume} {6}},\ \bibinfo {pages} {926} (\bibinfo {year} {2021})}\BibitemShut {NoStop}%
\bibitem [{\citenamefont {Melnikov}\ \emph {et~al.}(2015)\citenamefont {Melnikov}, \citenamefont {Shashkin}, \citenamefont {Dolgopolov}, \citenamefont {Huang}, \citenamefont {Liu},\ and\ \citenamefont {Kravchenko}}]{melnikov2015ultra}%
  \BibitemOpen
  \bibfield  {author} {\bibinfo {author} {\bibfnamefont {M.~Y.}\ \bibnamefont {Melnikov}}, \bibinfo {author} {\bibfnamefont {A.}~\bibnamefont {Shashkin}}, \bibinfo {author} {\bibfnamefont {V.}~\bibnamefont {Dolgopolov}}, \bibinfo {author} {\bibfnamefont {S.-H.}\ \bibnamefont {Huang}}, \bibinfo {author} {\bibfnamefont {C.}~\bibnamefont {Liu}}, \ and\ \bibinfo {author} {\bibfnamefont {S.}~\bibnamefont {Kravchenko}},\ }\href@noop {} {\bibfield  {journal} {\bibinfo  {journal} {Applied Physics Letters}\ }\textbf {\bibinfo {volume} {106}} (\bibinfo {year} {2015})}\BibitemShut {NoStop}%
\bibitem [{\citenamefont {Myronov}\ \emph {et~al.}(2023)\citenamefont {Myronov}, \citenamefont {Kycia}, \citenamefont {Waldron}, \citenamefont {Jiang}, \citenamefont {Barrios}, \citenamefont {Bogan}, \citenamefont {Coleridge},\ and\ \citenamefont {Studenikin}}]{myronov2023holes}%
  \BibitemOpen
  \bibfield  {author} {\bibinfo {author} {\bibfnamefont {M.}~\bibnamefont {Myronov}}, \bibinfo {author} {\bibfnamefont {J.}~\bibnamefont {Kycia}}, \bibinfo {author} {\bibfnamefont {P.}~\bibnamefont {Waldron}}, \bibinfo {author} {\bibfnamefont {W.}~\bibnamefont {Jiang}}, \bibinfo {author} {\bibfnamefont {P.}~\bibnamefont {Barrios}}, \bibinfo {author} {\bibfnamefont {A.}~\bibnamefont {Bogan}}, \bibinfo {author} {\bibfnamefont {P.}~\bibnamefont {Coleridge}}, \ and\ \bibinfo {author} {\bibfnamefont {S.}~\bibnamefont {Studenikin}},\ }\href@noop {} {\bibfield  {journal} {\bibinfo  {journal} {Small Science}\ }\textbf {\bibinfo {volume} {3}},\ \bibinfo {pages} {2200094} (\bibinfo {year} {2023})}\BibitemShut {NoStop}%
\bibitem [{\citenamefont {Ayan}\ \emph {et~al.}(2019)\citenamefont {Ayan}, \citenamefont {Turkay}, \citenamefont {Unlu}, \citenamefont {Naghinazhadahmadi}, \citenamefont {Oliaei}, \citenamefont {Boztug},\ and\ \citenamefont {Yerci}}]{ayan2019strain}%
  \BibitemOpen
  \bibfield  {author} {\bibinfo {author} {\bibfnamefont {A.}~\bibnamefont {Ayan}}, \bibinfo {author} {\bibfnamefont {D.}~\bibnamefont {Turkay}}, \bibinfo {author} {\bibfnamefont {B.}~\bibnamefont {Unlu}}, \bibinfo {author} {\bibfnamefont {P.}~\bibnamefont {Naghinazhadahmadi}}, \bibinfo {author} {\bibfnamefont {S.~N.~B.}\ \bibnamefont {Oliaei}}, \bibinfo {author} {\bibfnamefont {C.}~\bibnamefont {Boztug}}, \ and\ \bibinfo {author} {\bibfnamefont {S.}~\bibnamefont {Yerci}},\ }\href@noop {} {\bibfield  {journal} {\bibinfo  {journal} {Scientific Reports}\ }\textbf {\bibinfo {volume} {9}},\ \bibinfo {pages} {4963} (\bibinfo {year} {2019})}\BibitemShut {NoStop}%
\bibitem [{\citenamefont {Thurmond}(1953)}]{thurmond1953equilibrium}%
  \BibitemOpen
  \bibfield  {author} {\bibinfo {author} {\bibfnamefont {C.~D.}\ \bibnamefont {Thurmond}},\ }\href@noop {} {\bibfield  {journal} {\bibinfo  {journal} {The Journal of Physical Chemistry}\ }\textbf {\bibinfo {volume} {57}},\ \bibinfo {pages} {827} (\bibinfo {year} {1953})}\BibitemShut {NoStop}%
\bibitem [{\citenamefont {Li}\ \emph {et~al.}(2013)\citenamefont {Li}, \citenamefont {Cui}, \citenamefont {Wu}, \citenamefont {Tseng}, \citenamefont {Cheng},\ and\ \citenamefont {Chen}}]{li2013strain}%
  \BibitemOpen
  \bibfield  {author} {\bibinfo {author} {\bibfnamefont {H.}~\bibnamefont {Li}}, \bibinfo {author} {\bibfnamefont {Y.}~\bibnamefont {Cui}}, \bibinfo {author} {\bibfnamefont {K.}~\bibnamefont {Wu}}, \bibinfo {author} {\bibfnamefont {W.}~\bibnamefont {Tseng}}, \bibinfo {author} {\bibfnamefont {H.}~\bibnamefont {Cheng}}, \ and\ \bibinfo {author} {\bibfnamefont {H.}~\bibnamefont {Chen}},\ }\href@noop {} {\bibfield  {journal} {\bibinfo  {journal} {Applied Physics Letters}\ }\textbf {\bibinfo {volume} {102}} (\bibinfo {year} {2013})}\BibitemShut {NoStop}%
\bibitem [{\citenamefont {Cai}\ \emph {et~al.}(2022)\citenamefont {Cai}, \citenamefont {Qian}, \citenamefont {An}, \citenamefont {Lin}, \citenamefont {Wu}, \citenamefont {Ding}, \citenamefont {Huang}, \citenamefont {Chen}, \citenamefont {Wang},\ and\ \citenamefont {Li}}]{cai2022thickness}%
  \BibitemOpen
  \bibfield  {author} {\bibinfo {author} {\bibfnamefont {H.}~\bibnamefont {Cai}}, \bibinfo {author} {\bibfnamefont {K.}~\bibnamefont {Qian}}, \bibinfo {author} {\bibfnamefont {Y.}~\bibnamefont {An}}, \bibinfo {author} {\bibfnamefont {G.}~\bibnamefont {Lin}}, \bibinfo {author} {\bibfnamefont {S.}~\bibnamefont {Wu}}, \bibinfo {author} {\bibfnamefont {H.}~\bibnamefont {Ding}}, \bibinfo {author} {\bibfnamefont {W.}~\bibnamefont {Huang}}, \bibinfo {author} {\bibfnamefont {S.}~\bibnamefont {Chen}}, \bibinfo {author} {\bibfnamefont {J.}~\bibnamefont {Wang}}, \ and\ \bibinfo {author} {\bibfnamefont {C.}~\bibnamefont {Li}},\ }\href@noop {} {\bibfield  {journal} {\bibinfo  {journal} {Journal of Alloys and Compounds}\ }\textbf {\bibinfo {volume} {904}},\ \bibinfo {pages} {164068} (\bibinfo {year} {2022})}\BibitemShut {NoStop}%
\bibitem [{\citenamefont {Tsukamoto}\ \emph {et~al.}(2015)\citenamefont {Tsukamoto}, \citenamefont {Hirose}, \citenamefont {Kasamatsu}, \citenamefont {Mimura}, \citenamefont {Matsui},\ and\ \citenamefont {Suda}}]{tsukamoto2015investigation}%
  \BibitemOpen
  \bibfield  {author} {\bibinfo {author} {\bibfnamefont {T.}~\bibnamefont {Tsukamoto}}, \bibinfo {author} {\bibfnamefont {N.}~\bibnamefont {Hirose}}, \bibinfo {author} {\bibfnamefont {A.}~\bibnamefont {Kasamatsu}}, \bibinfo {author} {\bibfnamefont {T.}~\bibnamefont {Mimura}}, \bibinfo {author} {\bibfnamefont {T.}~\bibnamefont {Matsui}}, \ and\ \bibinfo {author} {\bibfnamefont {Y.}~\bibnamefont {Suda}},\ }\href@noop {} {\bibfield  {journal} {\bibinfo  {journal} {Applied Physics Letters}\ }\textbf {\bibinfo {volume} {106}} (\bibinfo {year} {2015})}\BibitemShut {NoStop}%
\bibitem [{\citenamefont {Li}\ \emph {et~al.}(2014)\citenamefont {Li}, \citenamefont {Chang}, \citenamefont {Chen}, \citenamefont {Cheng}, \citenamefont {Shi},\ and\ \citenamefont {Chen}}]{li2014characteristics}%
  \BibitemOpen
  \bibfield  {author} {\bibinfo {author} {\bibfnamefont {H.}~\bibnamefont {Li}}, \bibinfo {author} {\bibfnamefont {C.}~\bibnamefont {Chang}}, \bibinfo {author} {\bibfnamefont {T.}~\bibnamefont {Chen}}, \bibinfo {author} {\bibfnamefont {H.}~\bibnamefont {Cheng}}, \bibinfo {author} {\bibfnamefont {Z.}~\bibnamefont {Shi}}, \ and\ \bibinfo {author} {\bibfnamefont {H.}~\bibnamefont {Chen}},\ }\href@noop {} {\bibfield  {journal} {\bibinfo  {journal} {Applied Physics Letters}\ }\textbf {\bibinfo {volume} {105}} (\bibinfo {year} {2014})}\BibitemShut {NoStop}%
\bibitem [{\citenamefont {Wang}\ \emph {et~al.}(2020)\citenamefont {Wang}, \citenamefont {Zhang}, \citenamefont {Wu}, \citenamefont {Liu}, \citenamefont {Miao}, \citenamefont {Meng}, \citenamefont {Jiang},\ and\ \citenamefont {Hu}}]{wang2020effects}%
  \BibitemOpen
  \bibfield  {author} {\bibinfo {author} {\bibfnamefont {L.}~\bibnamefont {Wang}}, \bibinfo {author} {\bibfnamefont {Y.}~\bibnamefont {Zhang}}, \bibinfo {author} {\bibfnamefont {Y.}~\bibnamefont {Wu}}, \bibinfo {author} {\bibfnamefont {T.}~\bibnamefont {Liu}}, \bibinfo {author} {\bibfnamefont {Y.}~\bibnamefont {Miao}}, \bibinfo {author} {\bibfnamefont {L.}~\bibnamefont {Meng}}, \bibinfo {author} {\bibfnamefont {Z.}~\bibnamefont {Jiang}}, \ and\ \bibinfo {author} {\bibfnamefont {H.}~\bibnamefont {Hu}},\ }\href@noop {} {\bibfield  {journal} {\bibinfo  {journal} {IEEE Transactions on Electron Devices}\ }\textbf {\bibinfo {volume} {67}},\ \bibinfo {pages} {3229} (\bibinfo {year} {2020})}\BibitemShut {NoStop}%
\bibitem [{\citenamefont {Kong}\ \emph {et~al.}(2022)\citenamefont {Kong}, \citenamefont {Wang}, \citenamefont {Liang}, \citenamefont {Su}, \citenamefont {Xun}, \citenamefont {Miao}, \citenamefont {Gu}, \citenamefont {Li}, \citenamefont {Cao}, \citenamefont {Lin} \emph {et~al.}}]{kong2022growth}%
  \BibitemOpen
  \bibfield  {author} {\bibinfo {author} {\bibfnamefont {Z.}~\bibnamefont {Kong}}, \bibinfo {author} {\bibfnamefont {G.}~\bibnamefont {Wang}}, \bibinfo {author} {\bibfnamefont {R.}~\bibnamefont {Liang}}, \bibinfo {author} {\bibfnamefont {J.}~\bibnamefont {Su}}, \bibinfo {author} {\bibfnamefont {M.}~\bibnamefont {Xun}}, \bibinfo {author} {\bibfnamefont {Y.}~\bibnamefont {Miao}}, \bibinfo {author} {\bibfnamefont {S.}~\bibnamefont {Gu}}, \bibinfo {author} {\bibfnamefont {J.}~\bibnamefont {Li}}, \bibinfo {author} {\bibfnamefont {K.}~\bibnamefont {Cao}}, \bibinfo {author} {\bibfnamefont {H.}~\bibnamefont {Lin}},  \emph {et~al.},\ }\href@noop {} {\bibfield  {journal} {\bibinfo  {journal} {Nanomaterials}\ }\textbf {\bibinfo {volume} {12}},\ \bibinfo {pages} {981} (\bibinfo {year} {2022})}\BibitemShut {NoStop}%
\bibitem [{\citenamefont {Fischetti}\ and\ \citenamefont {Laux}(1996)}]{Fisch}%
  \BibitemOpen
  \bibfield  {author} {\bibinfo {author} {\bibfnamefont {M.~V.}\ \bibnamefont {Fischetti}}\ and\ \bibinfo {author} {\bibfnamefont {S.~E.}\ \bibnamefont {Laux}},\ }\href@noop {} {\bibfield  {journal} {\bibinfo  {journal} {Journal of Applied Physics}\ }\textbf {\bibinfo {volume} {80}},\ \bibinfo {pages} {2234} (\bibinfo {year} {1996})}\BibitemShut {NoStop}%
\bibitem [{\citenamefont {Prince}(1953)}]{prince}%
  \BibitemOpen
  \bibfield  {author} {\bibinfo {author} {\bibfnamefont {M.}~\bibnamefont {Prince}},\ }\href@noop {} {\bibfield  {journal} {\bibinfo  {journal} {Physical Review}\ }\textbf {\bibinfo {volume} {92}},\ \bibinfo {pages} {681} (\bibinfo {year} {1953})}\BibitemShut {NoStop}%
\bibitem [{\citenamefont {Rideau}\ \emph {et~al.}(2006)\citenamefont {Rideau}, \citenamefont {Feraille}, \citenamefont {Ciampolini}, \citenamefont {Minondo}, \citenamefont {Tavernier}, \citenamefont {Jaouen},\ and\ \citenamefont {Ghetti}}]{rideau}%
  \BibitemOpen
  \bibfield  {author} {\bibinfo {author} {\bibfnamefont {D.}~\bibnamefont {Rideau}}, \bibinfo {author} {\bibfnamefont {M.}~\bibnamefont {Feraille}}, \bibinfo {author} {\bibfnamefont {L.}~\bibnamefont {Ciampolini}}, \bibinfo {author} {\bibfnamefont {M.}~\bibnamefont {Minondo}}, \bibinfo {author} {\bibfnamefont {C.}~\bibnamefont {Tavernier}}, \bibinfo {author} {\bibfnamefont {H.}~\bibnamefont {Jaouen}}, \ and\ \bibinfo {author} {\bibfnamefont {A.}~\bibnamefont {Ghetti}},\ }\href@noop {} {\bibfield  {journal} {\bibinfo  {journal} {Physical Review B}\ }\textbf {\bibinfo {volume} {74}},\ \bibinfo {pages} {195208} (\bibinfo {year} {2006})}\BibitemShut {NoStop}%
\bibitem [{\citenamefont {El~Kurdi}\ \emph {et~al.}(2010{\natexlab{a}})\citenamefont {El~Kurdi}, \citenamefont {Fishman}, \citenamefont {Sauvage},\ and\ \citenamefont {Boucaud}}]{kurdi}%
  \BibitemOpen
  \bibfield  {author} {\bibinfo {author} {\bibfnamefont {M.}~\bibnamefont {El~Kurdi}}, \bibinfo {author} {\bibfnamefont {G.}~\bibnamefont {Fishman}}, \bibinfo {author} {\bibfnamefont {S.}~\bibnamefont {Sauvage}}, \ and\ \bibinfo {author} {\bibfnamefont {P.}~\bibnamefont {Boucaud}},\ }\href@noop {} {\bibfield  {journal} {\bibinfo  {journal} {Journal of Applied Physics}\ }\textbf {\bibinfo {volume} {107}} (\bibinfo {year} {2010}{\natexlab{a}})}\BibitemShut {NoStop}%
\bibitem [{\citenamefont {Kao}\ \emph {et~al.}(2014)\citenamefont {Kao}, \citenamefont {Verhulst}, \citenamefont {Van~de Put}, \citenamefont {Vandenberghe}, \citenamefont {Soree}, \citenamefont {Magnus},\ and\ \citenamefont {De~Meyer}}]{kao}%
  \BibitemOpen
  \bibfield  {author} {\bibinfo {author} {\bibfnamefont {K.-H.}\ \bibnamefont {Kao}}, \bibinfo {author} {\bibfnamefont {A.~S.}\ \bibnamefont {Verhulst}}, \bibinfo {author} {\bibfnamefont {M.}~\bibnamefont {Van~de Put}}, \bibinfo {author} {\bibfnamefont {W.~G.}\ \bibnamefont {Vandenberghe}}, \bibinfo {author} {\bibfnamefont {B.}~\bibnamefont {Soree}}, \bibinfo {author} {\bibfnamefont {W.}~\bibnamefont {Magnus}}, \ and\ \bibinfo {author} {\bibfnamefont {K.}~\bibnamefont {De~Meyer}},\ }\href@noop {} {\bibfield  {journal} {\bibinfo  {journal} {Journal of Applied Physics}\ }\textbf {\bibinfo {volume} {115}} (\bibinfo {year} {2014})}\BibitemShut {NoStop}%
\bibitem [{\citenamefont {Jiang}\ \emph {et~al.}(2014)\citenamefont {Jiang}, \citenamefont {Gallagher}, \citenamefont {Senaratne}, \citenamefont {Aoki}, \citenamefont {Mathews}, \citenamefont {Kouvetakis},\ and\ \citenamefont {Menendez}}]{jiang}%
  \BibitemOpen
  \bibfield  {author} {\bibinfo {author} {\bibfnamefont {L.}~\bibnamefont {Jiang}}, \bibinfo {author} {\bibfnamefont {J.}~\bibnamefont {Gallagher}}, \bibinfo {author} {\bibfnamefont {C.~L.}\ \bibnamefont {Senaratne}}, \bibinfo {author} {\bibfnamefont {T.}~\bibnamefont {Aoki}}, \bibinfo {author} {\bibfnamefont {J.}~\bibnamefont {Mathews}}, \bibinfo {author} {\bibfnamefont {J.}~\bibnamefont {Kouvetakis}}, \ and\ \bibinfo {author} {\bibfnamefont {J.}~\bibnamefont {Menendez}},\ }\href@noop {} {\bibfield  {journal} {\bibinfo  {journal} {Semiconductor Science and Technology}\ }\textbf {\bibinfo {volume} {29}},\ \bibinfo {pages} {115028} (\bibinfo {year} {2014})}\BibitemShut {NoStop}%
\bibitem [{\citenamefont {Ghetmiri}\ \emph {et~al.}(2014)\citenamefont {Ghetmiri}, \citenamefont {Du}, \citenamefont {Margetis}, \citenamefont {Mosleh}, \citenamefont {Cousar}, \citenamefont {Conley}, \citenamefont {Domulevicz}, \citenamefont {Nazzal}, \citenamefont {Sun}, \citenamefont {Soref} \emph {et~al.}}]{ghetmiri2014direct}%
  \BibitemOpen
  \bibfield  {author} {\bibinfo {author} {\bibfnamefont {S.~A.}\ \bibnamefont {Ghetmiri}}, \bibinfo {author} {\bibfnamefont {W.}~\bibnamefont {Du}}, \bibinfo {author} {\bibfnamefont {J.}~\bibnamefont {Margetis}}, \bibinfo {author} {\bibfnamefont {A.}~\bibnamefont {Mosleh}}, \bibinfo {author} {\bibfnamefont {L.}~\bibnamefont {Cousar}}, \bibinfo {author} {\bibfnamefont {B.~R.}\ \bibnamefont {Conley}}, \bibinfo {author} {\bibfnamefont {L.}~\bibnamefont {Domulevicz}}, \bibinfo {author} {\bibfnamefont {A.}~\bibnamefont {Nazzal}}, \bibinfo {author} {\bibfnamefont {G.}~\bibnamefont {Sun}}, \bibinfo {author} {\bibfnamefont {R.~A.}\ \bibnamefont {Soref}},  \emph {et~al.},\ }\href@noop {} {\bibfield  {journal} {\bibinfo  {journal} {Applied Physics Letters}\ }\textbf {\bibinfo {volume} {105}} (\bibinfo {year} {2014})}\BibitemShut {NoStop}%
\bibitem [{\citenamefont {Gallagher}\ \emph {et~al.}(2014)\citenamefont {Gallagher}, \citenamefont {Senaratne}, \citenamefont {Kouvetakis},\ and\ \citenamefont {Menendez}}]{gallagher2014compositional}%
  \BibitemOpen
  \bibfield  {author} {\bibinfo {author} {\bibfnamefont {J.}~\bibnamefont {Gallagher}}, \bibinfo {author} {\bibfnamefont {C.~L.}\ \bibnamefont {Senaratne}}, \bibinfo {author} {\bibfnamefont {J.}~\bibnamefont {Kouvetakis}}, \ and\ \bibinfo {author} {\bibfnamefont {J.}~\bibnamefont {Menendez}},\ }\href@noop {} {\bibfield  {journal} {\bibinfo  {journal} {Applied Physics Letters}\ }\textbf {\bibinfo {volume} {105}} (\bibinfo {year} {2014})}\BibitemShut {NoStop}%
\bibitem [{\citenamefont {Clavel}\ \emph {et~al.}(2022)\citenamefont {Clavel}, \citenamefont {Murphy-Armando}, \citenamefont {Xie}, \citenamefont {Henry}, \citenamefont {Kuhn}, \citenamefont {Bodnar}, \citenamefont {Khodaparast}, \citenamefont {Smirnov}, \citenamefont {Heremans},\ and\ \citenamefont {Hudait}}]{clavelfma}%
  \BibitemOpen
  \bibfield  {author} {\bibinfo {author} {\bibfnamefont {M.~B.}\ \bibnamefont {Clavel}}, \bibinfo {author} {\bibfnamefont {F.}~\bibnamefont {Murphy-Armando}}, \bibinfo {author} {\bibfnamefont {Y.}~\bibnamefont {Xie}}, \bibinfo {author} {\bibfnamefont {K.}~\bibnamefont {Henry}}, \bibinfo {author} {\bibfnamefont {M.}~\bibnamefont {Kuhn}}, \bibinfo {author} {\bibfnamefont {R.~J.}\ \bibnamefont {Bodnar}}, \bibinfo {author} {\bibfnamefont {G.}~\bibnamefont {Khodaparast}}, \bibinfo {author} {\bibfnamefont {D.}~\bibnamefont {Smirnov}}, \bibinfo {author} {\bibfnamefont {J.}~\bibnamefont {Heremans}}, \ and\ \bibinfo {author} {\bibfnamefont {M.~K.}\ \bibnamefont {Hudait}},\ }\href@noop {} {\bibfield  {journal} {\bibinfo  {journal} {Physical Review Applied}\ }\textbf {\bibinfo {volume} {18}},\ \bibinfo {pages} {064083} (\bibinfo {year} {2022})}\BibitemShut {NoStop}%
\bibitem [{\citenamefont {Soref}\ \emph {et~al.}(2006)\citenamefont {Soref}, \citenamefont {Kouvetakis},\ and\ \citenamefont {Menendez}}]{soref2006advances}%
  \BibitemOpen
  \bibfield  {author} {\bibinfo {author} {\bibfnamefont {R.}~\bibnamefont {Soref}}, \bibinfo {author} {\bibfnamefont {J.}~\bibnamefont {Kouvetakis}}, \ and\ \bibinfo {author} {\bibfnamefont {J.}~\bibnamefont {Menendez}},\ }\href@noop {} {\bibfield  {journal} {\bibinfo  {journal} {MRS Online Proceedings Library (OPL)}\ }\textbf {\bibinfo {volume} {958}},\ \bibinfo {pages} {0958} (\bibinfo {year} {2006})}\BibitemShut {NoStop}%
\bibitem [{\citenamefont {Saladukha}\ \emph {et~al.}(2018)\citenamefont {Saladukha}, \citenamefont {Clavel}, \citenamefont {Murphy-Armando}, \citenamefont {Greene-Diniz}, \citenamefont {Gr{\"u}ning}, \citenamefont {Hudait},\ and\ \citenamefont {Ochalski}}]{saladukha2018direct}%
  \BibitemOpen
  \bibfield  {author} {\bibinfo {author} {\bibfnamefont {D.}~\bibnamefont {Saladukha}}, \bibinfo {author} {\bibfnamefont {M.}~\bibnamefont {Clavel}}, \bibinfo {author} {\bibfnamefont {F.}~\bibnamefont {Murphy-Armando}}, \bibinfo {author} {\bibfnamefont {G.}~\bibnamefont {Greene-Diniz}}, \bibinfo {author} {\bibfnamefont {M.}~\bibnamefont {Gr{\"u}ning}}, \bibinfo {author} {\bibfnamefont {M.}~\bibnamefont {Hudait}}, \ and\ \bibinfo {author} {\bibfnamefont {T.~J.}\ \bibnamefont {Ochalski}},\ }\href@noop {} {\bibfield  {journal} {\bibinfo  {journal} {Physical Review B}\ }\textbf {\bibinfo {volume} {97}},\ \bibinfo {pages} {195304} (\bibinfo {year} {2018})}\BibitemShut {NoStop}%
\bibitem [{\citenamefont {Pavarelli}\ \emph {et~al.}(2013)\citenamefont {Pavarelli}, \citenamefont {Ochalski}, \citenamefont {Murphy-Armando}, \citenamefont {Huo}, \citenamefont {Schmidt}, \citenamefont {Huyet},\ and\ \citenamefont {Harris}}]{pavarelli2013optical}%
  \BibitemOpen
  \bibfield  {author} {\bibinfo {author} {\bibfnamefont {N.}~\bibnamefont {Pavarelli}}, \bibinfo {author} {\bibfnamefont {T.~J.}\ \bibnamefont {Ochalski}}, \bibinfo {author} {\bibfnamefont {F.}~\bibnamefont {Murphy-Armando}}, \bibinfo {author} {\bibfnamefont {Y.}~\bibnamefont {Huo}}, \bibinfo {author} {\bibfnamefont {M.}~\bibnamefont {Schmidt}}, \bibinfo {author} {\bibfnamefont {G.}~\bibnamefont {Huyet}}, \ and\ \bibinfo {author} {\bibfnamefont {J.}~\bibnamefont {Harris}},\ }\href@noop {} {\bibfield  {journal} {\bibinfo  {journal} {Physical Review Letters}\ }\textbf {\bibinfo {volume} {110}},\ \bibinfo {pages} {177404} (\bibinfo {year} {2013})}\BibitemShut {NoStop}%
\bibitem [{\citenamefont {Murphy-Armando}\ and\ \citenamefont {Fahy}(2006)}]{fma06}%
  \BibitemOpen
  \bibfield  {author} {\bibinfo {author} {\bibfnamefont {F.}~\bibnamefont {Murphy-Armando}}\ and\ \bibinfo {author} {\bibfnamefont {S.}~\bibnamefont {Fahy}},\ }\href@noop {} {\bibfield  {journal} {\bibinfo  {journal} {Physical review letters}\ }\textbf {\bibinfo {volume} {97}},\ \bibinfo {pages} {096606} (\bibinfo {year} {2006})}\BibitemShut {NoStop}%
\bibitem [{\citenamefont {Liu}\ \emph {et~al.}(2023)\citenamefont {Liu}, \citenamefont {Junk}, \citenamefont {Han}, \citenamefont {Yang}, \citenamefont {Bae}, \citenamefont {Frauenrath}, \citenamefont {Hartmann}, \citenamefont {Ikonic}, \citenamefont {B{\"a}rwolf}, \citenamefont {Mai} \emph {et~al.}}]{liu2023vertical}%
  \BibitemOpen
  \bibfield  {author} {\bibinfo {author} {\bibfnamefont {M.}~\bibnamefont {Liu}}, \bibinfo {author} {\bibfnamefont {Y.}~\bibnamefont {Junk}}, \bibinfo {author} {\bibfnamefont {Y.}~\bibnamefont {Han}}, \bibinfo {author} {\bibfnamefont {D.}~\bibnamefont {Yang}}, \bibinfo {author} {\bibfnamefont {J.~H.}\ \bibnamefont {Bae}}, \bibinfo {author} {\bibfnamefont {M.}~\bibnamefont {Frauenrath}}, \bibinfo {author} {\bibfnamefont {J.-M.}\ \bibnamefont {Hartmann}}, \bibinfo {author} {\bibfnamefont {Z.}~\bibnamefont {Ikonic}}, \bibinfo {author} {\bibfnamefont {F.}~\bibnamefont {B{\"a}rwolf}}, \bibinfo {author} {\bibfnamefont {A.}~\bibnamefont {Mai}},  \emph {et~al.},\ }\href@noop {} {\bibfield  {journal} {\bibinfo  {journal} {Communications Engineering}\ }\textbf {\bibinfo {volume} {2}},\ \bibinfo {pages} {7} (\bibinfo {year} {2023})}\BibitemShut {NoStop}%
\bibitem [{\citenamefont {Murphy-Armando}(2019)}]{fmathermo}%
  \BibitemOpen
  \bibfield  {author} {\bibinfo {author} {\bibfnamefont {F.}~\bibnamefont {Murphy-Armando}},\ }\href@noop {} {\bibfield  {journal} {\bibinfo  {journal} {Journal of Applied Physics}\ }\textbf {\bibinfo {volume} {126}} (\bibinfo {year} {2019})}\BibitemShut {NoStop}%
\bibitem [{\citenamefont {Jacoboni}\ and\ \citenamefont {Reggiani}(1983)}]{jac}%
  \BibitemOpen
  \bibfield  {author} {\bibinfo {author} {\bibfnamefont {C.}~\bibnamefont {Jacoboni}}\ and\ \bibinfo {author} {\bibfnamefont {L.}~\bibnamefont {Reggiani}},\ }\href@noop {} {\bibfield  {journal} {\bibinfo  {journal} {Reviews of modern Physics}\ }\textbf {\bibinfo {volume} {55}},\ \bibinfo {pages} {645} (\bibinfo {year} {1983})}\BibitemShut {NoStop}%
\bibitem [{\citenamefont {Paige}(1964)}]{paige1964electrical}%
  \BibitemOpen
  \bibfield  {author} {\bibinfo {author} {\bibfnamefont {E.}~\bibnamefont {Paige}},\ }\href@noop {} {\bibfield  {journal} {\bibinfo  {journal} {Semiconductors}\ }\textbf {\bibinfo {volume} {8}},\ \bibinfo {pages} {202} (\bibinfo {year} {1964})}\BibitemShut {NoStop}%
\bibitem [{\citenamefont {Ashcroft}\ and\ \citenamefont {Mermin}(1976)}]{ashcroft1976solid}%
  \BibitemOpen
  \bibfield  {author} {\bibinfo {author} {\bibfnamefont {N.~W.}\ \bibnamefont {Ashcroft}}\ and\ \bibinfo {author} {\bibfnamefont {N.~D.}\ \bibnamefont {Mermin}},\ }\href@noop {} {\bibfield  {journal} {\bibinfo  {journal} {Thomson Learning Inc}\ } (\bibinfo {year} {1976})}\BibitemShut {NoStop}%
\bibitem [{\citenamefont {Hedin}(1965)}]{Hedin1965}%
  \BibitemOpen
  \bibfield  {author} {\bibinfo {author} {\bibfnamefont {L.}~\bibnamefont {Hedin}},\ }\href {\doibase 10.1103/physrev.139.a796} {\bibfield  {journal} {\bibinfo  {journal} {Phys. Rev.}\ }\textbf {\bibinfo {volume} {139}},\ \bibinfo {pages} {A796} (\bibinfo {year} {1965})}\BibitemShut {NoStop}%
\bibitem [{\citenamefont {Hybertsen}\ and\ \citenamefont {Louie}(1986)}]{hybertsen1986electron}%
  \BibitemOpen
  \bibfield  {author} {\bibinfo {author} {\bibfnamefont {M.~S.}\ \bibnamefont {Hybertsen}}\ and\ \bibinfo {author} {\bibfnamefont {S.~G.}\ \bibnamefont {Louie}},\ }\href@noop {} {\bibfield  {journal} {\bibinfo  {journal} {Physical Review B}\ }\textbf {\bibinfo {volume} {34}},\ \bibinfo {pages} {5390} (\bibinfo {year} {1986})}\BibitemShut {NoStop}%
\bibitem [{\citenamefont {Gonze}\ \emph {et~al.}(2009)\citenamefont {Gonze}, \citenamefont {Amadon}, \citenamefont {Anglade}, \citenamefont {Beuken}, \citenamefont {Bottin}, \citenamefont {Boulanger}, \citenamefont {Bruneval}, \citenamefont {Caliste}, \citenamefont {Caracas}, \citenamefont {C{\^o}t{\'e}} \emph {et~al.}}]{ab}%
  \BibitemOpen
  \bibfield  {author} {\bibinfo {author} {\bibfnamefont {X.}~\bibnamefont {Gonze}}, \bibinfo {author} {\bibfnamefont {B.}~\bibnamefont {Amadon}}, \bibinfo {author} {\bibfnamefont {P.-M.}\ \bibnamefont {Anglade}}, \bibinfo {author} {\bibfnamefont {J.-M.}\ \bibnamefont {Beuken}}, \bibinfo {author} {\bibfnamefont {F.}~\bibnamefont {Bottin}}, \bibinfo {author} {\bibfnamefont {P.}~\bibnamefont {Boulanger}}, \bibinfo {author} {\bibfnamefont {F.}~\bibnamefont {Bruneval}}, \bibinfo {author} {\bibfnamefont {D.}~\bibnamefont {Caliste}}, \bibinfo {author} {\bibfnamefont {R.}~\bibnamefont {Caracas}}, \bibinfo {author} {\bibfnamefont {M.}~\bibnamefont {C{\^o}t{\'e}}},  \emph {et~al.},\ }\href@noop {} {\bibfield  {journal} {\bibinfo  {journal} {Computer Physics Communications}\ }\textbf {\bibinfo {volume} {180}},\ \bibinfo {pages} {2582} (\bibinfo {year} {2009})}\BibitemShut {NoStop}%
\bibitem [{\citenamefont {Aulbur}\ \emph {et~al.}(2000)\citenamefont {Aulbur}, \citenamefont {St{\"a}dele},\ and\ \citenamefont {G{\"o}rling}}]{GW}%
  \BibitemOpen
  \bibfield  {author} {\bibinfo {author} {\bibfnamefont {W.~G.}\ \bibnamefont {Aulbur}}, \bibinfo {author} {\bibfnamefont {M.}~\bibnamefont {St{\"a}dele}}, \ and\ \bibinfo {author} {\bibfnamefont {A.}~\bibnamefont {G{\"o}rling}},\ }\href@noop {} {\bibfield  {journal} {\bibinfo  {journal} {Physical Review B}\ }\textbf {\bibinfo {volume} {62}},\ \bibinfo {pages} {7121} (\bibinfo {year} {2000})}\BibitemShut {NoStop}%
\bibitem [{\citenamefont {Varshni}(1967)}]{varshni1967temperature}%
  \BibitemOpen
  \bibfield  {author} {\bibinfo {author} {\bibfnamefont {Y.~P.}\ \bibnamefont {Varshni}},\ }\href@noop {} {\bibfield  {journal} {\bibinfo  {journal} {physica}\ }\textbf {\bibinfo {volume} {34}},\ \bibinfo {pages} {149} (\bibinfo {year} {1967})}\BibitemShut {NoStop}%
\bibitem [{\citenamefont {Precker}\ and\ \citenamefont {da~Silva}(2002)}]{precker2002experimental}%
  \BibitemOpen
  \bibfield  {author} {\bibinfo {author} {\bibfnamefont {J.~W.}\ \bibnamefont {Precker}}\ and\ \bibinfo {author} {\bibfnamefont {M.~A.}\ \bibnamefont {da~Silva}},\ }\href@noop {} {\bibfield  {journal} {\bibinfo  {journal} {American Journal of Physics}\ }\textbf {\bibinfo {volume} {70}},\ \bibinfo {pages} {1150} (\bibinfo {year} {2002})}\BibitemShut {NoStop}%
\bibitem [{\citenamefont {El~Kurdi}\ \emph {et~al.}(2010{\natexlab{b}})\citenamefont {El~Kurdi}, \citenamefont {Bertin}, \citenamefont {Martincic}, \citenamefont {De~Kersauson}, \citenamefont {Fishman}, \citenamefont {Sauvage}, \citenamefont {Bosseboeuf},\ and\ \citenamefont {Boucaud}}]{el2010control}%
  \BibitemOpen
  \bibfield  {author} {\bibinfo {author} {\bibfnamefont {M.}~\bibnamefont {El~Kurdi}}, \bibinfo {author} {\bibfnamefont {H.}~\bibnamefont {Bertin}}, \bibinfo {author} {\bibfnamefont {E.}~\bibnamefont {Martincic}}, \bibinfo {author} {\bibfnamefont {M.}~\bibnamefont {De~Kersauson}}, \bibinfo {author} {\bibfnamefont {G.}~\bibnamefont {Fishman}}, \bibinfo {author} {\bibfnamefont {S.}~\bibnamefont {Sauvage}}, \bibinfo {author} {\bibfnamefont {A.}~\bibnamefont {Bosseboeuf}}, \ and\ \bibinfo {author} {\bibfnamefont {P.}~\bibnamefont {Boucaud}},\ }\href@noop {} {\bibfield  {journal} {\bibinfo  {journal} {Applied Physics Letters}\ }\textbf {\bibinfo {volume} {96}} (\bibinfo {year} {2010}{\natexlab{b}})}\BibitemShut {NoStop}%
\bibitem [{\citenamefont {Dexter}\ \emph {et~al.}(1956)\citenamefont {Dexter}, \citenamefont {Zeiger},\ and\ \citenamefont {Lax}}]{dexter1956cyclotron}%
  \BibitemOpen
  \bibfield  {author} {\bibinfo {author} {\bibfnamefont {R.}~\bibnamefont {Dexter}}, \bibinfo {author} {\bibfnamefont {H.}~\bibnamefont {Zeiger}}, \ and\ \bibinfo {author} {\bibfnamefont {B.}~\bibnamefont {Lax}},\ }\href@noop {} {\bibfield  {journal} {\bibinfo  {journal} {Physical Review}\ }\textbf {\bibinfo {volume} {104}},\ \bibinfo {pages} {637} (\bibinfo {year} {1956})}\BibitemShut {NoStop}%
\bibitem [{\citenamefont {Murphy-Armando}\ and\ \citenamefont {Fahy}(2008)}]{fma08}%
  \BibitemOpen
  \bibfield  {author} {\bibinfo {author} {\bibfnamefont {F.}~\bibnamefont {Murphy-Armando}}\ and\ \bibinfo {author} {\bibfnamefont {S.}~\bibnamefont {Fahy}},\ }\href@noop {} {\bibfield  {journal} {\bibinfo  {journal} {Physical Review B}\ }\textbf {\bibinfo {volume} {78}},\ \bibinfo {pages} {035202} (\bibinfo {year} {2008})}\BibitemShut {NoStop}%
\bibitem [{\citenamefont {Herring}\ and\ \citenamefont {Vogt}(1956)}]{hv}%
  \BibitemOpen
  \bibfield  {author} {\bibinfo {author} {\bibfnamefont {C.}~\bibnamefont {Herring}}\ and\ \bibinfo {author} {\bibfnamefont {E.}~\bibnamefont {Vogt}},\ }\href@noop {} {\bibfield  {journal} {\bibinfo  {journal} {Physical review}\ }\textbf {\bibinfo {volume} {101}},\ \bibinfo {pages} {944} (\bibinfo {year} {1956})}\BibitemShut {NoStop}%
\bibitem [{\citenamefont {Harmin}(1982)}]{harmin1982theory}%
  \BibitemOpen
  \bibfield  {author} {\bibinfo {author} {\bibfnamefont {D.~A.}\ \bibnamefont {Harmin}},\ }\href@noop {} {\bibfield  {journal} {\bibinfo  {journal} {Physical review A}\ }\textbf {\bibinfo {volume} {26}},\ \bibinfo {pages} {2656} (\bibinfo {year} {1982})}\BibitemShut {NoStop}%
\bibitem [{\citenamefont {Paul}(2008)}]{paul20088}%
  \BibitemOpen
  \bibfield  {author} {\bibinfo {author} {\bibfnamefont {D.~J.}\ \bibnamefont {Paul}},\ }\href@noop {} {\bibfield  {journal} {\bibinfo  {journal} {Physical Review B}\ }\textbf {\bibinfo {volume} {77}},\ \bibinfo {pages} {155323} (\bibinfo {year} {2008})}\BibitemShut {NoStop}%
\bibitem [{\citenamefont {Hartwigsen}\ \emph {et~al.}(1998)\citenamefont {Hartwigsen}, \citenamefont {G{\oe}decker},\ and\ \citenamefont {Hutter}}]{hgh}%
  \BibitemOpen
  \bibfield  {author} {\bibinfo {author} {\bibfnamefont {C.}~\bibnamefont {Hartwigsen}}, \bibinfo {author} {\bibfnamefont {S.}~\bibnamefont {G{\oe}decker}}, \ and\ \bibinfo {author} {\bibfnamefont {J.}~\bibnamefont {Hutter}},\ }\href@noop {} {\bibfield  {journal} {\bibinfo  {journal} {Physical Review B}\ }\textbf {\bibinfo {volume} {58}},\ \bibinfo {pages} {3641} (\bibinfo {year} {1998})}\BibitemShut {NoStop}%
\end{thebibliography}

%

\end{document}